\documentclass{article}
\usepackage[super,compress]{cite}
\usepackage{mathrsfs,amsfonts,amsmath}
\usepackage{graphicx}
\usepackage{subfigure}
\usepackage{bm}
\newcommand{\Tr}{\textrm{Tr}}
\newcommand{\Lag}{\mathscr{L}}
\newcommand{\refcite}{\citen}
\DeclareMathOperator{\A}{\mathbf{A}}
\DeclareMathOperator{\B}{\mathbf{B}}

\begin{document}

\markboth{Carlos Naya}{Neutron Stars within the Skyrme Model}


\title{{\bf Neutron Stars within the Skyrme Model}}

\author{Carlos Naya \\ \small {\it INFN, Sezione di Lecce}\\
\small {\it Via per Arnesano, C.P. 193 I-73100 Lecce, Italy} \\ \small {\it carlos.naya@le.infn.it}}
\date{}


\maketitle


\begin{abstract}
The Skyrme model is a low energy effective field theory of strong interactions where nuclei and baryons appear as collective excitations of pionic degrees of freedom. In the last years, there has been a revival of Skyrme's ideas and new related models, some of them with BPS bounds (topological lower energy bounds), have been proposed. It is the aim of this paper to review how they can be applied to the study of neutron stars allowing for a description by means of topological solitons. We will focus on different aspects as the equation of state or the mass-radius relation, where we find that high maximal masses are supported.
\end{abstract}




\section{Introduction}

Proposed by the British physicist T. H. R. Skyrme in the sixties of last century, \cite{Skyrme1961,Skyrme1962} the Skyrme model appears as an effective field theory of strong interactions at low energies with pions being the fundamental degrees of freedom. Nuclei and baryons then emerge as topological solitons, \cite{TopologicalSolitons} {\it i.e.}, collective excitations of the pionic fields whose stability comes from the topology of the base and target space. They are particle-like solutions of non-linear field equations with a conserved quantity, the topological charge, which can be identified with the baryon number $B$.

This novel idea of Skyrmions to describe the non-perturbative regime of Quantum Chromodynamics (QCD) received more recognition when it was discovered that in the QCD limit of a large number of colors an effective field theory of mesons arises. \cite{Witten1983} As a consequence, interesting studies of the nucleon and its quantization \cite{Adkins1983,Adkins1984} together with the deuteron \cite{Braaten1986,Braaten1989} and some light nuclei \cite{Carson1991a,Carson1991b,Walhout1991} followed, with important qualitative results for such a simple theory. At the same time, there were some attempts for a description of neutron stars within the theory by using the simplest possible ansatz. \cite{Glendenning1988,Bizon1992} Unfortunately, this straightforward approach didn't do well and the model seemed to have been forgotten in the following years.

With the new millennium, there has been a revival of Skyrme's original ideas bringing an intense activity to the field. The application of the rational map ansatz to Skyrmions \cite{Houghton1998} has allowed a systematic and more complete description of light nuclei and their rotational spectra and quantization \cite{Battye2001,Battye2002,Battye2005,Battye2006,Manko2007,Battye2009} but also the extension to the study of other aspects as isospinning Skyrmions, \cite{Battye2015} Skyrmion-Skyrmion scattering,\cite{Foster2015a,Foster2015b} and more recently, their vibrational modes. \cite{Gudnason2018a}  At the same time, the success of generalized integrability \cite{Alvarez1998} (see Ref.~\refcite{Alvarez2009} for a review and applications) boosted the search of a submodel with infinite conserved currents and symmetries \cite{Ferreira2001} for the development of exact methods for the Skyrme model.

This renewed interest has been accompanied with more results agreeing qualitatively with the field of nuclear physics but also with important quantitative insights as for instance in the case of the well-known isotope of Carbon-12. \cite{Lau2014} Nevertheless, despite its importance as a bridge between the fundamental theory of QCD and nuclear physics, the original theory proposed by Skyrme has also some important drawbacks which can be summarized in the physically too-large nuclear binding energies and the lack of cluster structures.

In the last years, some attempts focused on the improvement of the binding energy issue have been carried out by the study of related Skyrme-like theories. Here there exist two main approaches: the lightly bound model of Ref.~\refcite{Gillard2015} based on a variation of the potential present in the Lagrangian by the inclusion of an additional term, or the path of Refs.~\refcite{Paul2010,Adam2010a,Adam2010b} consisting in a modification to achieve a BPS theory. \cite{Bogomolny1976,Prasad1975} Note that BPS implies there is a lower topological bound for the energy which is linear with the topological charge (baryon number). Then, saturating the bound would mean that the corresponding BPS configurations would have identically zero binding energies from the classical point of view. Also these two BPS models are quite different from each other. The one from Ref.~\refcite{Adam2010a} is an extreme case of a generalized Skyrme model (see \ref{BPS-Model} for more information) which arrives at low biding energies by taking into account small but necessary additional contributions \cite{Adam2013a,Adam2013b} and which will play an important role for the forthcoming description of neutron stars. On the other hand, the theory in Ref.~\refcite{Paul2010} corresponds to the Skyrme model coupled to an infinite tower of vector mesons and has the outstanding property that the inclusion of the lightest mesons not only reduces the binding energies in a considerable way but also solves the last-standing problem of clustering in the Skyrme model. \cite{Naya2018a,Naya2018b}

Concentrating now on the description of neutron stars from the perspective of topological solitons, there have also been important developments and improvements coming together with the advances in the description of nuclei. Hence, after the already mentioned first attempts with the simplest ansatz, the implementation of the rational map was also extended and refined to the study of neutron stars, \cite{Piette2007,Nelmes2011} whilst the fact that for large baryon number the true minimal configuration corresponds to a crystal lattice of $B=4$ Skyrmions (alpha-particles) \cite{Castillejo1989} led to a crystal-like characterization of these compact objects.\cite{Nelmes2012} More recently, the BPS Skyrme model of Ref.~\refcite{Adam2010a} has brought a perfect fluid description to the field,\cite{Adam2015a,Adam2015b} in consonance with what is believed to be the state of matter in the neutron star core. On the other hand, some studies with a half-Skyrmion phase, which is relevant to dense nuclear matter and compact stars, have been presented. \cite{Rho2011,Rho2013a,Rho2013b,Rho2018}

In addition, although out of the scope of this review, some works relating black holes and the Skyrme model have been also carried out where Skyrmionic hair has been studied. \cite{Sawado2004,Shiiki2005a,Shiiki2005b,Brihaye2006,Perapechka2017,Adam2016a,Gudnason2016,Gudnason2018}

In the following, we will review these different attempts and descriptions of the important objects which neutron stars represent, showing how the Skyrme model does not only restrict itself to the study of light nuclei but emerges as a unified field theory of nuclear matter. In section \ref{SkyrmeModels} we will introduce the Skyrme model in its original form but paying also attention to the BPS version of interest for neutron stars. The corresponding implications and the study of neutron stars as topological solitons will be presented in section \ref{NeutronStars} to end with some conclusions and important remarks in section \ref{Conclusions}.


\section{Skyrme Models} \label{SkyrmeModels}

The Skyrme model is an effective field theory of strong interactions which in its standard version is given by the Lagrangian
\begin{equation}\label{StandardLag}
\Lag_{Sk} = -\frac{f_\pi^2}{4} \Tr \left( L_\mu L^\mu \right) + \frac{1}{32 e^2} \Tr \left( [L_\mu,L_\nu] [L^\mu,L^\nu] \right) - \frac{m_\pi^2 f_\pi^2}{8} \Tr(1-U) ,
\end{equation}

\noindent with $f_\pi$, $e$ and $m_\pi$ the pion decay constant, the Skyrme coupling and the pion mass, respectively. $L_\mu = U^\dagger \partial_\mu U$ are the $\mathfrak{su}$(2)-valued and left invariant Maurer-Cartan currents and $U \in SU(2)$ the Skyrme field. Due to the constant boundary value of $U$ when $x \rightarrow \infty$, the base space can be compactified to the three-sphere, $\mathbb{S}^3$, and thanks to the fact that the manifold SU(2) is also isomorphic to  $\mathbb{S}^3$, the Skyrme field appears as a map between two three-spheres. As a consequence of its topological character, there is a conserved topological number which may be identified with the baryon number and reads
\begin{equation}\label{ChargeB}
B = \frac{1}{24 \pi^2} \int d^3x \, \epsilon^{i j k} \, \Tr \left(L_i L_j L_k \right) .
\end{equation}

The first term in the Lagrangian corresponds to the non-linear sigma model and is quadratic in first derivatives, encoding the kinetic energy of pions. The second contribution, known as Skyrme term and quartic in first derivatives, appears to circumvent Derrick's theorem so stable solutions can exist. \cite{Derrick1964} Finally, we also have a potential giving mass to pions (sometimes the mass term is neglected for simplicity and we can talk about the massless Skyrme model instead of the massive one). To see that pions are actually the fundamental degrees of freedom, we need to recall that $U$ is an SU(2)-valued field where the pion field triplet $(\pi_1,\pi_2,\pi_3)$, can be encoded as follows
\begin{equation}
U= \left( \begin{array}{ccc}
\sigma + i \pi_3 & \quad & i \pi_1 + \pi_2 \\
i \pi_1 - \pi_2 & \quad & \sigma - i \pi_3
\end{array} \right) ,
\end{equation}

\noindent with $\sigma$ an auxiliary field imposing the constraint $\sigma^2 + \pi_1^2 + \pi_2^2 + \pi_3^2 =1$. Considering small perturbations of the Skyrme field around its vacuum expectation value, the behavior of each term in the Lagrangian in dimensionless units is
\begin{equation}
\Lag_2 =  -\frac{1}{2} \Tr \left(U^\dagger \partial_\mu U U^\dagger \partial^\mu U \right) \sim \frac{1}{2} \partial_\mu \vec{\pi} \cdot \partial^\mu \vec{\pi} + O(\pi^4),
\end{equation}
\begin{equation}
\Lag_4 = \frac{1}{16} \Tr \left([U^\dagger \partial_\mu U, U^\dagger \partial_\nu U] [U^\dagger \partial^\mu U, U^\dagger \partial^\nu U] \right) \sim O(\pi^4),
\end{equation}
\begin{equation}
\Lag_0 = m_\pi^2 \Tr(U-1) \sim - \frac{1}{2} m_\pi^2 \vec{\pi}^2 + O(\pi^4).
\end{equation}

\noindent Thus, adding them up we get
\begin{equation}
\Lag_\pi = \frac{1}{2} \partial_\mu \vec{\pi} \cdot \partial^\mu \vec{\pi} - \frac{1}{2} m_\pi^2 \vec{\pi}^2 + O(\pi^4),
\end{equation}

\noindent which is the expected Lagrangian for a pionic field.

The fact that the Skyrme term is of quartic order in $\pi$ is a welcome outcome since we know that its motivation comes from Derrick's theorem. Important to note is also that the model has only three parameters with one being the pion mass. A common practice is to set the remaining values to $f_\pi = 129 \; {\rm MeV}$ and $e=5.45$, which were found by Adkins, Nappi and Witten in Ref.~\refcite{Adkins1983} by quantizing the $B=1$ Skyrmion and fitting the parameters to the mass values of the nucleon and the delta baryon\footnote{Where not stated otherwise, these are the values we will assume.} (different values have been also proposed, see for instance Ref.~\refcite{Adam2016b}).

Indeed, looking for solutions, the model can be easily solved only for the nucleon, {\it i.e.}, the $B=1$ soliton. In this case, the right configuration is spherically symmetric and given by the hedgehog ansatz
\begin{equation}\label{hedgehog}
U=\exp(i f(r) \, \vec{\tau} \cdot \hat r) ,
\end{equation}

\noindent where $\vec{\tau} = (\tau_1,\tau_2, \tau_3)$ are the Pauli matrices and the field $f(r)$, known as the profile function, takes the boundary values $f(r=0)=\pi$, $f(r=\infty)=0$. Hence, the energy density minimization is reduced to a one-dimensional problem and $f(r)$ can be computed numerically.

Fortunately, it has been shown that for baryon number greater than one, there is a good and reliable approximation, the rational map. Used before in the study of BPS monopoles, it has ended playing an important role for Skyrmions too.\cite{Houghton1998} Working with spherical coordinates in $\mathbb{R}^3$, the ansatz for the Skyrme field reads
\begin{equation}\label{RM_ansatz}
U = \exp(i f(r) \, \vec{\tau} \cdot \hat n_R) ,
\end{equation}

\noindent where again the profile function $f(r)$ is subject to the boundary conditions $f(r=0)=\pi$, $f(r=\infty)=0$, and $\hat{n}_R$ is the unit vector
\begin{equation}
\hat{n}_R = \frac{1}{1+|R|^2} \left( 2 \Re(R), 2 \Im(R), 1-|R|^2 \right).
\end{equation}

\noindent Here, $R$ is a rational function of the stereographic coordinate $z=\tan(\theta/2) \, e^{i \phi}$ and the baryon number of the soliton is given by the degree of the rational map $R$. To find an approximate solution within the ansatz for a fixed value of $B$, one has to look for a map $R$ minimizing the integral 
\begin{equation}\label{I_RM}
\mathcal I = \frac{1}{4 \pi} \int \left( \frac{1+|z|^2}{1+|R|^2} \left| \frac{dR}{dz}\right| \right)^4 \frac{2 i \,dz \,d\bar z}{(1+|z|^2)^2},
\end{equation}

\noindent and next minimize the energy functional with respect to the profile function $f(r)$. In addition, the integral $\mathcal I$ can be approximated by $\mathcal I \approx 1.28 B^2$ for large baryon number, \cite{Battye2002} reducing again the problem in this regime to one dimension.

Finally, let us note that since we are working with an effective theory, we could think about adding more terms to Eq. (\ref{StandardLag}). In that case, to write down the most general Lagrangian with at most second order time derivatives to allow for a Hamiltonian formulation, the only possible Lorentz-invariant contribution is
\begin{equation}
\Lag_6 = - \lambda^2 \pi^2 \mathcal{B}_\mu \mathcal{B}^\mu .
\end{equation}

\noindent Here, $\lambda$ is a coupling constant while 
\begin{equation}\label{Bmu}
\mathcal{B}^\mu = \frac{1}{24 \pi^2} \Tr(\epsilon^{\mu \nu \rho \sigma} U^\dagger \partial_\nu U U^\dagger \partial_\rho U U^\dagger \partial_\sigma U)
\end{equation}

\noindent is the baryon current. Bear in mind that $\mathcal{B}^0$ corresponds to the baryon density whose integral is nothing but the conserved charge $B$ as in Eq. (\ref{ChargeB}). This $\Lag_6$ will be of key importance for the construction of the BPS model we are interested in for the description of neutron stars.


\subsection{The BPS Skyrme model} \label{BPS-Model}

The BPS Skyrme model proposed by C. Adam, J. Sanchez-Guillen and A. Wereszczynski in Refs.~\refcite{Adam2010a,Adam2010b} may be thought of as an extreme case of the most general possible Lagrangian where only the contribution from the square of the topological current, $\Lag_6$, and a potential are kept, namely,
\begin{equation}
\Lag_{\rm BPS} = \Lag_6 + \Lag_0 = -\lambda^2 \pi^2 \mathcal{B}_\mu \mathcal{B}^\mu - \mu^2 \, \mathcal U,
\end{equation}

\noindent where $\mathcal{U}$ is a potential which does not have to be necessarily the usual pion mass potential of Eq. (\ref{StandardLag}), and $\lambda$ and $\mu$ are the two coupling constants of the theory. This may be thought as the limit of an infinite pion mass, and since $\Lag_6$ is sextic in first derivatives, there is no need for a Skyrme term to stabilize the soliton anymore. The main issue with this theory is the absence of pion dynamics; however, it should be considered as an idealization giving the main contribution of a near-BPS theory defined by the Lagrangian
\begin{equation}
\Lag = \varepsilon \, \Lag_{Sk} + \Lag_{\rm BPS} \, ,
\end{equation}

\noindent with $\varepsilon$ a small parameter and $\Lag_{Sk}$ the Lagrangian of Eq. (\ref{StandardLag}). 

Therefore, the study of the BPS part of this generalized Skyrme model is important since it is expected that some properties of nuclei, baryons and nuclear matter will be governed by it, while the effect of $\Lag_{Sk}$ will be dominant near the vacuum. Furthermore, the sextic term is related to the coupling of the omega mesons, \cite{Adam2015c} which is known to be the main interaction channel at high density/pressure and appears as an effect to take into consideration when dealing with the compact objects which neutron stars represent.

On the other hand, this limiting case presents some significant features which are welcome and will help with the applications, like its solvability or the energy-momentum tensor of a perfect fluid. Let us start with the property giving name to the model, the BPS bound. As previously mentioned, it is a topological lower bound on the energy which does not depend on a particular solution. To see it, we can assume the following parametrization of the Skyrme field,
\begin{equation}
U = \exp( i \xi \, \hat n \cdot \vec \tau) = \cos \xi + i \sin \xi \, \hat n \cdot \vec \tau,
\end{equation}

\noindent with $\xi=\xi(\vec r)$ the profile function and $\vec{\tau}$ the Pauli matrices. The unit vector $\hat n$ can be written in terms of a complex field $u$ by means of the stereographic projection
\begin{equation}
\hat n = \frac{1}{1+|u|^2} \left(u + \bar u,-i (u - \bar u), 1 - |u|^2 \right),
\end{equation}

\noindent and considering that the potential only depends on the Skyrme field through $\Tr(U)$, {\it i.e.}, $\mathcal U = \mathcal U(\Tr U) = \mathcal U(\xi)$, the BPS Lagrangian reads
\begin{equation}\label{BPSLag}
\Lag_{\rm BPS} = \frac{\lambda^2 \sin^4 \xi}{(1 + |u|^2)^4} (\epsilon^{\mu \nu \rho \sigma} \partial_\nu \xi \, \partial_\rho u \, \partial_\sigma \bar u)^2 - \mu^2 \, \mathcal U(\xi).
\end{equation}

A useful trick to find the BPS bound is to write the static energy functional and complete the square. Hence, from
\begin{equation}
E_{\rm BPS} = \int \left( \frac{\lambda^2 \sin^4 \xi}{(1 + |u|^2)^4} (\epsilon^{m n l} i \partial_m \xi \, \partial_n u \, \partial_l \bar u)^2 + \mu^2 \, \mathcal U(\xi) \right) d^3 x ,
\end{equation}

\noindent we obtain the topological bound
\begin{equation}\label{BPSBound}
E_{\rm BPS} \geq 2 \pi^2 \lambda \mu \, \langle \sqrt \mathcal U \rangle_{\mathbb{S}^3} |B|, \qquad \langle \sqrt \mathcal U \rangle_{\mathbb{S}^3} = \frac{1}{2 \pi^2} \int_{\mathbb{S}^3} \sqrt{\mathcal U (\xi)} \, d\Omega,
\end{equation}

\noindent where $\langle \sqrt \mathcal U \rangle_{\mathbb{S}^3}$ is the average of the square root of the potential over the target space. Note that the integral does not depend on any particular solution of the model. In addition, the requirement to fulfil the bound implies the BPS equation
\begin{equation}
\frac{\lambda \sin^2 \xi}{(1+|u|^2)^2} \epsilon^{m n l} i \partial_m \xi \, \partial_n u \, \partial_l \bar u = \mp \mu \sqrt \mathcal U.
\end{equation}

\noindent This is an additional simplification since we are going from the equations of motion, which are second order partial differential equations (PDE), to a first order PDE. 

Back to the energy functional, any coordinate transformation on the base space leaving the volume form, $d^3x$, invariant is a symmetry transformation. Hence, the volume preserving diffeomorphisms (VPDs) on the base space $\mathbb{S}^3$ are symmetries of the model, but these are exactly the symmetries of a perfect fluid. On the other hand, we also have an infinite number of target space symmetries. The contribution $\Lag_6$ is the squared pullback of
\begin{equation}
d\Omega= - 2 i \frac{sin^2 \xi}{(1 + |u|^2)^2}\, d\xi \, du \, d\bar u,
\end{equation}

\noindent the volume form on the target space $SU(2) \sim \mathbb{S}^3$ (note it is the same volume form appearing in Eq. (\ref{BPSBound})). Therefore, it has all VPD on the target space as symmetries. However, this is broken down to the area preserving diffeomorphisms on the two-sphere defined by the complex field $u$ due to the presence of the potential $\Lag_0$. All in all, the BPS Lagrangian is invariant under the changes $\xi \rightarrow \xi$, $u \rightarrow \tilde u(u,\bar u, \xi)$, such that 
\begin{equation}
(1+| \tilde u|^2)^{-2} d\xi \, d \tilde u \, d  \bar{\tilde{u}} = (1+|u|^2)^{-2} \, d\xi \, du \, d \bar u.
\end{equation}

Finally it is also worth commenting on the energy-momentum tensor, which is that of a perfect fluid. In fact, for static configurations we arrive at \cite{Adam2014}
\begin{equation}
T^{\mu \nu}=(p + \rho) \, u^\mu u^\nu - p \, \eta^{\mu \nu},
\end{equation}

\noindent with $\eta^{\mu \nu} = {\rm diag}(1,-1,-1,-1)$, the four-velocity $u^\mu = (1,0,0,0)$, and the energy density and pressure
\begin{equation}
\rho = \lambda^2 \pi^4 \mathcal B_0^2 + \mu^2 \mathcal U, \qquad p = \lambda^2 \pi^4 \mathcal B_0^2 - \mu^2 \mathcal U.
\end{equation}

From the conservation of the energy-momentum tensor, we obtain that the pressure takes a constant value $P$, together with the equation (called constant pressure equation)
\begin{equation}\label{ConstP}
\frac{|B| \lambda}{2 r^2} \sin^2 \xi \, \xi_r = - \mu \sqrt{\mathcal U + \tilde P} \qquad (P = \mu^2 \tilde P),
\end{equation}

\noindent where we have also assumed $\xi = \xi(r)$. This is a first integral of the static field equation which for $P=0$ reduces to the BPS one. A nice feature of the system is that without needing to know a specific solution we can integrate (\ref{ConstP}) to obtain the equation of state (EoS) of the BPS Skyrme model, that is to say,
\begin{equation}\label{BPS_EoS}
V(P)= 2 \pi \frac{\lambda}{\mu} |B| \int_0^\pi \frac{sin^2 \xi d \xi}{\sqrt{\mathcal U + \tilde P}} = \pi^2 \frac{\lambda}{\mu} |B| \left \langle \frac{1}{\sqrt{\mathcal U + \tilde P}} \right \rangle.
\end{equation}

\noindent And by using the constant pressure equation with the energy,
\begin{equation}\label{BPS_E}
E(P)= 2 \pi \lambda \mu |B| \int_0^\pi \frac{2 \mathcal U + \tilde P}{\sqrt{\mathcal U + \tilde P}} \sin^2 \xi d\xi = \pi^2  \lambda \mu |B| \left \langle \frac{2 \mathcal U + \tilde P}{\sqrt{\mathcal U + \tilde P}} \right \rangle,
\end{equation}

\noindent where we have expressed the quantities as averages over the target space:
\begin{equation}
\langle F \rangle \equiv \frac{1}{2 \pi^2} \int_{\mathbb S^3} F \, d\Omega.
\end{equation}

Let us stress again that both (\ref{BPS_EoS}) and (\ref{BPS_E}) are general expressions known without solving the system. This is of capital importance since we can define an average energy density
\begin{equation}\label{MF_rho}
\bar \rho=\frac{E}{V},
\end{equation}

\noindent which will allow us to have a two-fold approach for the study on neutron stars within the BPS Skyrme model. Therefore, besides an exact calculation where the back-reaction of gravity is included, Eq. (\ref{MF_rho}) will let us perform the usual mean field (MF) approximation.

If the reader is interested, an exhaustive review of the BPS Skyrme model and its applications can be found in Refs.~\refcite{NayaThesis,AdamChapter}.


\section{Neutron Stars as Solitons}\label{NeutronStars}

After this brief introduction to the relevant Skyrme models, it is time to focus on their application to neutron stars. These very compact objects under extreme conditions present the densest matter with gravity being essential for their understanding (one can see Ref.~\refcite{Lattimer2004} for an accesible review). For this reason, they are good natural ``laboratories" to check our knowledge of nuclear matter at high densities. They are of key importance for the study of the EoS of nuclear matter and can rule out some models by the analysis of the mass-radius diagram where there are excluded regions which any reasonable EoS should not reach (for instance, see Fig. 3 in Ref.~\refcite{Lattimer2012}, or Ref.~\refcite{Lattimer2017} for a more recent discussion).

As one might expect, these are massive stars with small radii. In fact, one difficulty some of the typical EoSs of nuclear physics have is to predict masses high enough to match some of the recent discoveries, as the PSR J2215+5135 binary of Ref.~\refcite{Linares2018} with a mass of about 2.3 $M_\odot$.

In the following, we are going to review how the Skyrme model approaches their description at zero temperature with different success rate. For this task, we will follow the same path the Skyrme model has walked when dealing with nuclei and which we have already sketched in section \ref{SkyrmeModels}: we will start from the hedgehog ansatz to the rational map and finishing with BPS Skyrme model's perfect fluid behavior, going through some novel approaches as the Skyrme crystal.


\subsection{Neutron stars in the standard Skyrme model}

The first thing to do is to couple the Skyrme field to gravity. This is given by the Einstein-Skyrme action
\begin{equation}\label{S-gravity}
S = \int \sqrt{-g} \left( \Lag_{M} - \frac{R}{16 \pi G} \right) d^4 x,
\end{equation}

\noindent with $R$ the Ricci scalar, $G$ the gravitational constant and $\Lag_M$ the matter part of the action, which corresponds either to $\Lag_{Sk}$ or $\Lag_{\rm BPS}$ depending on our model of interest but with covariant derivatives instead of the usual ones from flat space. In the remainder of this subsection we will use the former Lagrangian of Eq. (\ref{StandardLag}).

Similarly, the baryon number in curved space, is given by
\begin{equation}
B = \int \, \mathcal B^0 \, d^3x,
\end{equation}

\noindent where $\mathcal B^0$ is again the baryon density corresponding to the zero component of Eq. (\ref{Bmu}) but with covariant derivatives.

Due to the fact that we will work with very large baryon number and gravity prefers spherical symmetry, we will consider the following spherically symmetry metric
\begin{equation}\label{metricPiette}
ds^2 = -A^2(r) \left( 1 - \frac{2 m(r)}{r} \right) dt^2 + \left( 1 - \frac{2 m(r)}{r} \right)^{-1} dr^2 + r^2 (d\theta^2 + \sin^2 \theta d\phi^2),
\end{equation}

\noindent where $A(r)$ and $m(r)$ are the radial metric functions to determine. Hence, the Ricci scalar we will work with reads
\begin{equation}
R =\frac{-2}{A r^2} (-A'' r^2 - 2 A' r + 2 A'' r m + A' m + 3 A' r m' + A r m'' + 2 A m').
\end{equation}

Finally, there is only one more thing missing, the Skyrme field. From now on until the end of the section, we are going to study how different ans\"atze affect the description of  neutron stars (if the reader is interested in an exhaustive discussion of the standard Skyrme model and neutron stars, please have a look at Ref.~\refcite{NelmesThesis}).

\subsubsection{The hedgehog ansatz}

As with the study of nucleons and baryons, the first attempts to couple Skyrmions to gravity consisted in using the simplest known configuration, the hedgehog ansatz of Eq. (\ref{hedgehog}).\cite{Glendenning1988,Bizon1992} To extend its domain from the nucleon to solitons with topological charge larger than one in order to allow the huge baryon number $B$ of neutron stars, the radial profile function $f(r)$ is now constrained by the boundary conditions $f(r = 0) = B \pi$ and $f(r = \infty) = 0$.

Hence, the resulting Hamiltonian for massless pions is given by
\begin{equation}
H = \int A \left(-\frac{m'}{G} + \mathcal E \right) dr + \frac{m(\infty)}{G},
\end{equation}

\noindent where $m(\infty)$ is the asymptotic value of the mass and
\begin{equation}
\mathcal E = 4 \pi \left[ f_\pi^2 \left(\frac{1}{2} r^2 S f'^2 + \sin^2 f \right)+ \frac{\sin^2 f}{ e^2 r^2} \left( \frac{1}{2} \sin^2 f+ r^2 S f'^2 \right) \right],
\end{equation}

\noindent with $S(r) = 1- \frac{2 m(r)}{r}$. On the other hand, the baryon number can be calculated as
\begin{equation}
B= - \frac{2}{\pi } \int \sin^2 f f' dr.
\end{equation}

We can now derive the corresponding equations of motion. Introducing for convenience the dimensionless variables $x=e f_\pi r$ and $\mu(x)=e f_\pi m(r)$, the equations for the fields $f(x)$, $\mu(x)$ and $A(x)$ are
\begin{equation}\label{Hedgehog_Eqmu}
\mu' = \alpha \left[ \frac{1}{2} x^2 S f'^2 + \sin^2 f \left(1 + S f'^2 + \frac{\sin^2 f}{2 x^2} \right) \right],
\end{equation}
\begin{equation}\label{Hedgehog_Eqf}
[(x^2 + 2 \sin^2 f) A S f']' = A \sin(2 f) \left(1 + S f'^2 + \frac{\sin^2 f}{x^2} \right),
\end{equation}
\begin{equation}\label{Hedgehog_EqA}
A' = \alpha \left( x + \frac{2}{x} \sin^2 f \right) A f'^2,
\end{equation}

\noindent where the dimensionless effective coupling constant $\alpha = 4 \pi G f_\pi^2$ has been defined. Indeed, it is clear now that the the system does not depend on $f_\pi$ and $e$ anymore but only on the $\alpha$ parameter. Then, when $\alpha = 0$ we recover either the flat space limit of the Skyrme model if $\mu(x) = 0$, or the Schwarzschild background for $\mu(x) = constant >0$. In addition, the metric field $A$ can be trivially decoupled from the system by using Eq. (\ref{Hedgehog_EqA}).

An initial insight can be achieved by considering an ansatz for the profile function too.\cite{Glendenning1988} Although it already gives an idea about the main issue with this approach, a more reliable way of studying the problem is the one followed by P. Bizon and T. Chmaj in Ref.~\refcite{Bizon1992}. To solve the PDE system defined by Eq. (\ref{Hedgehog_Eqmu}), (\ref{Hedgehog_Eqf}) and (\ref{Hedgehog_EqA}), a shooting from the center of the star is used, where the free parameter to vary is the derivative of the profile function at $x=0$, {\it i.e.}, $f'(0)$.

The first thing to note is that for fixed baryon number there is an upper bound for the effective coupling, $\alpha_{\rm crit}^B$, and beyond this value the configuration collapses. It decreases with the square of $B$ and the proportionality constant is approximately the critical value of the $B=1$ gravitating Skyrmion (first studied in Ref.~\refcite{Droz1991}); therefore, we can consider $\alpha_{\rm crit}^B \approx 0.040378/B^2$.

For $\alpha < \alpha_{\rm crit}^B$, there are two branches of solutions as can be seen in Fig. 1 from Ref.~\refcite{Bizon1992} for $B=1$. To understand this behavior, we should consider the limit $\alpha \rightarrow 0$. From its definition, this implies either the gravity constant $G \rightarrow 0$ or the pion decay constant $f_\pi \rightarrow 0$. The first case corresponds to the lower branch, and since gravity is decoupled, we are left with the usual flat space solutions; this branch is stable. On the other hand, the upper branch tends to  solutions of the static spherically symmetric magnetic Einstein-Yang-Mills equations studied by Bartnik and Mckinnon in Ref.~\refcite{Bartnik1988}; the branch is unstable.

When analyzing the solutions with high baryon number an important concern arises, they are unstable under decay into $B=1$ gravitating Skyrmions. Although it might seem a big problem of the theory, we should bear in mind that the same situation appears in flat space, the hedgehog configuration is not a good ansatz for baryon number bigger than one. Then, as in the description of nuclei, we should seek a better answer, the rational map.

\subsubsection{Rational maps}

The next logical implementation for a better description of neutron stars is what we have already learnt with the study of nuclei and consider the rational map ansatz \cite{Houghton1998} introduced in Eq. (\ref{RM_ansatz}). Keeping the spherically symmetric metric of Eq. (\ref{metricPiette}), and slightly modifying the definition of dimensionless variables by a factor 1/2 to be $x=ef_\pi r/2$ and $\mu(x) = e f_\pi m(r)/2$, the Hamiltonian corresponding to the rational map approximation with massless pions reads \cite{Piette2007}
\begin{eqnarray} \label{H_RM}
&H = &\frac{16 \pi f_\pi}{e} \left[ \int_0^\infty A(x) \left(\frac{1}{2} S(x) f'^2(x) \, x^2 + B \sin^2 f(x) \left(1 + S(x) f'^2(x) \right) \right. \right. \nonumber \\
&& + \left. \left. \frac{\mathcal I \sin^4 f(x)}{2 x^2} - \frac{\mu'(x)}{\alpha} \right) dx + \frac{\mu(\infty)}{\alpha} \right].
\end{eqnarray}

On the other hand, the new equations of motion are now
\begin{eqnarray}
&\mu(x)' = &\alpha \left( \frac{1}{2} x^2 S(x) f'^2 (x) + B \sin^2 f(x)  \right. \nonumber \\
&& + B S(x) f'^2 (x) \sin^2 f(x) + \left. \frac{\mathcal I \sin^4 f(x)}{2x^2} \right),
\end{eqnarray}
\begin{eqnarray}
&f(x)'' = & \frac{1}{S(x) V(x)} \left[ \sin \left(2 f(x)\right) \left( B + B S(x) f'^2(x) + \frac{\mathcal I \sin^2 f(x)}{x^2} \right) \right. \nonumber \\
&& - \left. \frac{\alpha S(x) f'^3 (x) V^2(x)}{x} - S'(x) f'(x) V(x) - S(x) f'(x) V'(x) \right],
\end{eqnarray}
\begin{equation}
A(x)' = \alpha A(x) f'^2(x) \left( x + 2 B \, \frac{\sin^2 f(x)}{x} \right),
\end{equation}

\noindent with $\alpha = \pi f_\pi^2 G/2$ being again a dimensionless coupling constant and
\begin{equation}
V(x) = x^2 + 2 B \sin^2 f(x).
\end{equation}

Remember the integral $\mathcal I$ is given by Eq. (\ref{I_RM}) and minimized by a rational map $R$. Hence, one has to minimize it first to then use the value of $\mathcal I$ as a parameter of the field equations.  Fortunately, for large baryon number there is an approximation which can be directly used, \cite{Battye2002} $\mathcal I \approx 1.28 B^2$, simplifying the procedure to solve the system.

\begin{figure}[th]
\centerline{\includegraphics[width=12cm]{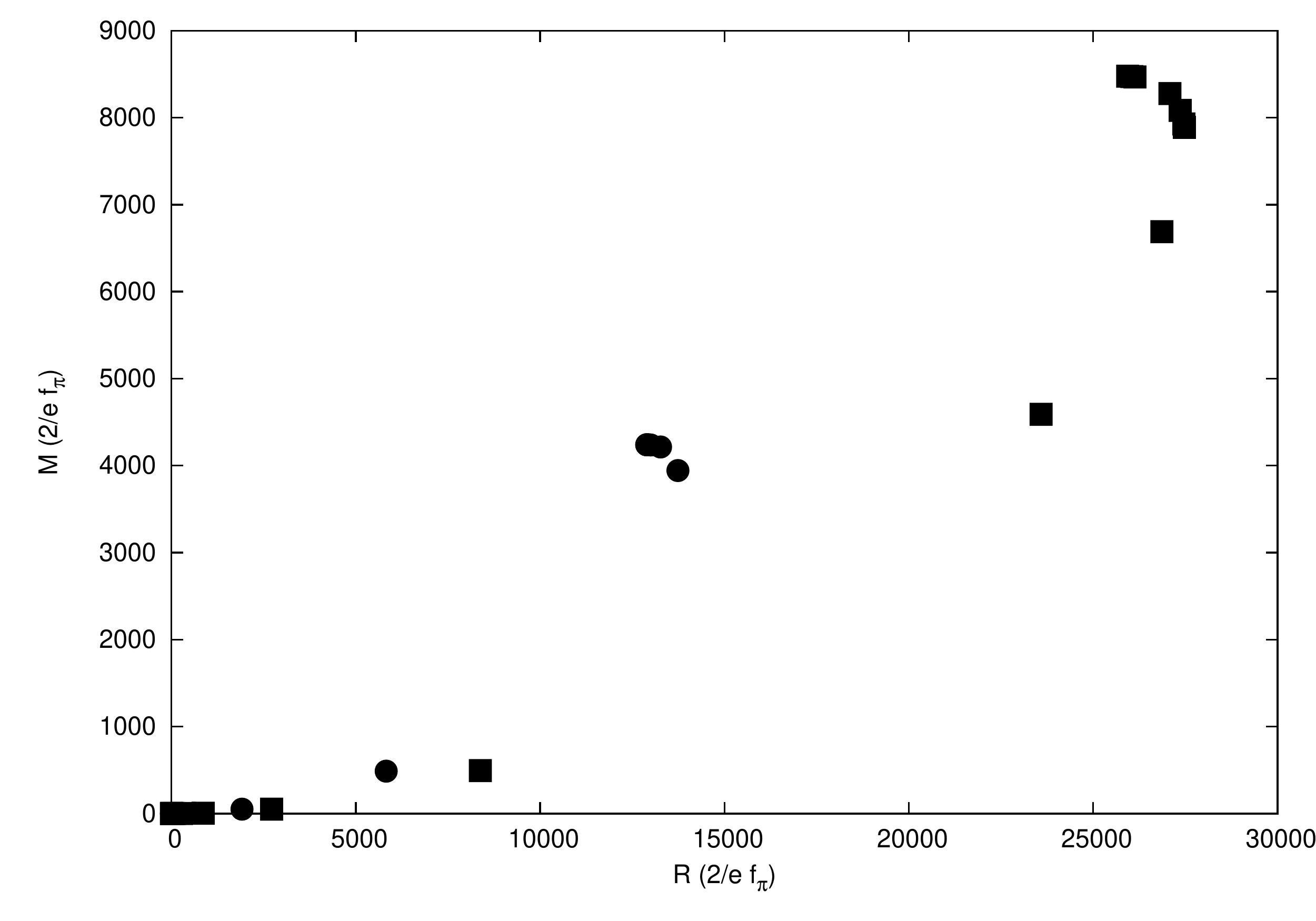}}
\caption{Neutron star mass as a function of the radius under the rational map approximation for $\alpha = 10^{-6}$. Square points correspond to the usual monolayer ansatz whilst circles represent the two-layer configuration. Plot obtained from data in Ref.~\refcite{Piette2007}.}\label{MuvsR-RM}
\end{figure}

We find a similar situation as with the hedgehog configuration. For given $B$, there is a critical value of the effective coupling $\alpha$ decreasing as $1/B^{1/2}$ and with two different branches where the lower is stable. However, rational map solutions are more bound for bigger $B$, so the problem of single-particle decay can be overcome for Skyrmionic neutron stars based on rational maps. Regarding the radius, it grows with the square root of the baryon number although the proportionality constant decreases a little when increasing it (an effect of the gravitational interaction), while near the maximal value of $B$, the radius is reduced by adding more baryons (an effect of the gravitational pull) as can be seen in Fig. \ref{MuvsR-RM}.

The relevant fields $f(x)$, $\mu(x)$ and $A(x)$ are step-like functions of the radial distance, giving rise to hollow shells (see Fig. 3 in Ref.~\refcite{Piette2007}). When increasing $B$, the shell radius (distance before the functions change) increases while the shell width (distance where the change happens) tends to a constant size, leading to nearly spherical structures for high values.

Another approach is to allow for different layers, which requires the boundary conditions for the profile function to be $f(r = 0) = N \pi$, $f(r = \infty) = 0$. Then, the topological charge is $N$ times the degree of the rational map and the functions appear as $N$ equal-size steps stacked together. In this case, the radius has decreased in a considerable way with respect to single-layer solutions, and when $B$ is large, double-layer solutions are energetically preferred although with a lowering of the neutron star mass (see Fig. \ref{MuvsR-RM} for $N=2$). For fixed $B$, they present a maximum baryon number which is almost twice that from single ones (about $1.95 \times 10^9$). However,  it is still far from $10^{57}$, the order of the physical value for neutron stars.

Handling physical high baryon numbers in this manner is extremely difficult since the layers would need to be of the order of $10^{17}$. To deal with this in a better way, we can follow Kopeliovich's work in Refs.~\refcite{Kopeliovich2001} and \refcite{Kopeliovich2002} and approximate the fields by piecewise linear profiles. This ramp-profile approximation \cite{Piette2007} will give us the chance to describe neutron stars with the desired baryon number by considering a large number of layers all of them carrying the same fraction of $B$. Although there seems to be a maximum number of shells before the solution ceases to exist, neutron stars with the desired baryon number and radius comparable to those in nature are found.

Despite this long desired success, this multilayer ansatz can be further improved by extending the ramp-profile approximation so both the baryon number and the width of the shell may vary along the neutron star. \cite{Nelmes2011} Following the same notation used by Nelmes and Piette, we introduce the number of baryons per shell, $b_i$, and the shell width $W_i$. Hence, the radial baryon density at each shell (please do not confuse with the spatial components of the topological current $\mathcal{B}^i$) is given by
\begin{equation}
B_i = \frac{b_i}{W_i} = - f'(x) \, \frac{b_i}{\pi},
\end{equation}

\noindent where we have used the approximation for the derivative of the profile function $-f'(x) \approx \pi/W_i$, which takes into account the fact that $f(x)$ varies from $n\pi$ to $(n-1)\pi$ within a given shell. Considering an extension to the continuum of the radial charge density at each shell, $b_i \rightarrow b(x)$, the total baryon number reads
\begin{equation}
B = - \int_0^R b(x) \, \frac{f'(x)}{\pi} \, dx,
\end{equation}

\noindent with
\begin{equation}
R = \sum_{i=1}^N W_i,
\end{equation}

\noindent the total radius ($W=R/N$ is the average shell width). 

Similarly, the Hamiltonian in Eq. (\ref{H_RM}) can be transformed into
\begin{eqnarray}
& H = & \frac{2}{e f_\pi G} \left[ \int_0^R \left[ -A(x) \mu'(x) + \alpha A(x) \left( x^2 S(x) f'^2(x) + b(x) \left(1 + S(x) f'^2(x)\right) \right. \right. \right. \nonumber \\
&& + \left. \left. \left. 1.28 \, b^2(x) \frac{3}{8 x^2} + 2 \mu_\pi^2 x^2 \right) \right]dx + \mu(\infty) \right],
\end{eqnarray}

\noindent where the previous approximation of the integral $\mathcal I$ at high baryon density translates into $\mathcal I \approx 1.28 \, b^2(x)$. In addition, we have used that, for an arbitrary function $G(x)$ slightly varying along the layer, the integral
\begin{equation}
\int G(x) \sin^p f(x) dx \approx \int G(x_0) \sin^p f(x) dx,
\end{equation}

\noindent together with
\begin{equation}
\int_{x_0 - W_i/2}^{x_0 + W_i/2} \sin^p f(x) dx = \frac{W_i}{\pi} \int_0^\pi \sin^p t \, dt,
\end{equation}

\noindent being $x_0$ the position of the shell. Note that, for convenience, we have also included the contribution from the pion mass in its dimensionless form $\mu_\pi = 2 m_\pi/(e f_\pi)$. Nevertheless, we will assume zero pion mass if not stated otherwise.

Finally, defining $q(x) = b(x)/x^2$ and introducing the scaling $f(x) = N g(x)$, $\mu(x) = N \nu(x)$ and  $x = N y$, we arrive at 
\begin{eqnarray}
&H =& \frac{2 N^3}{e f_\pi G} \int_0^W \left[ - A(y) \frac{\nu'(y)}{N^2} + \alpha y^2 A(y) \left[ \left(1-\frac{2 \nu(y)}{y} \right) g^2(y) \left(1 + q(y) \right) \right. \right. \nonumber \\
&& + \left. \left. q(y) + 1.28 \, q^2(y) \frac{3}{8} + 2 \mu_\pi^2 \right] \right] dy + \frac{2 N \nu(\infty)}{e f_\pi G}.
\end{eqnarray}

\begin{figure}[t]
\centerline{\includegraphics[width=12cm]{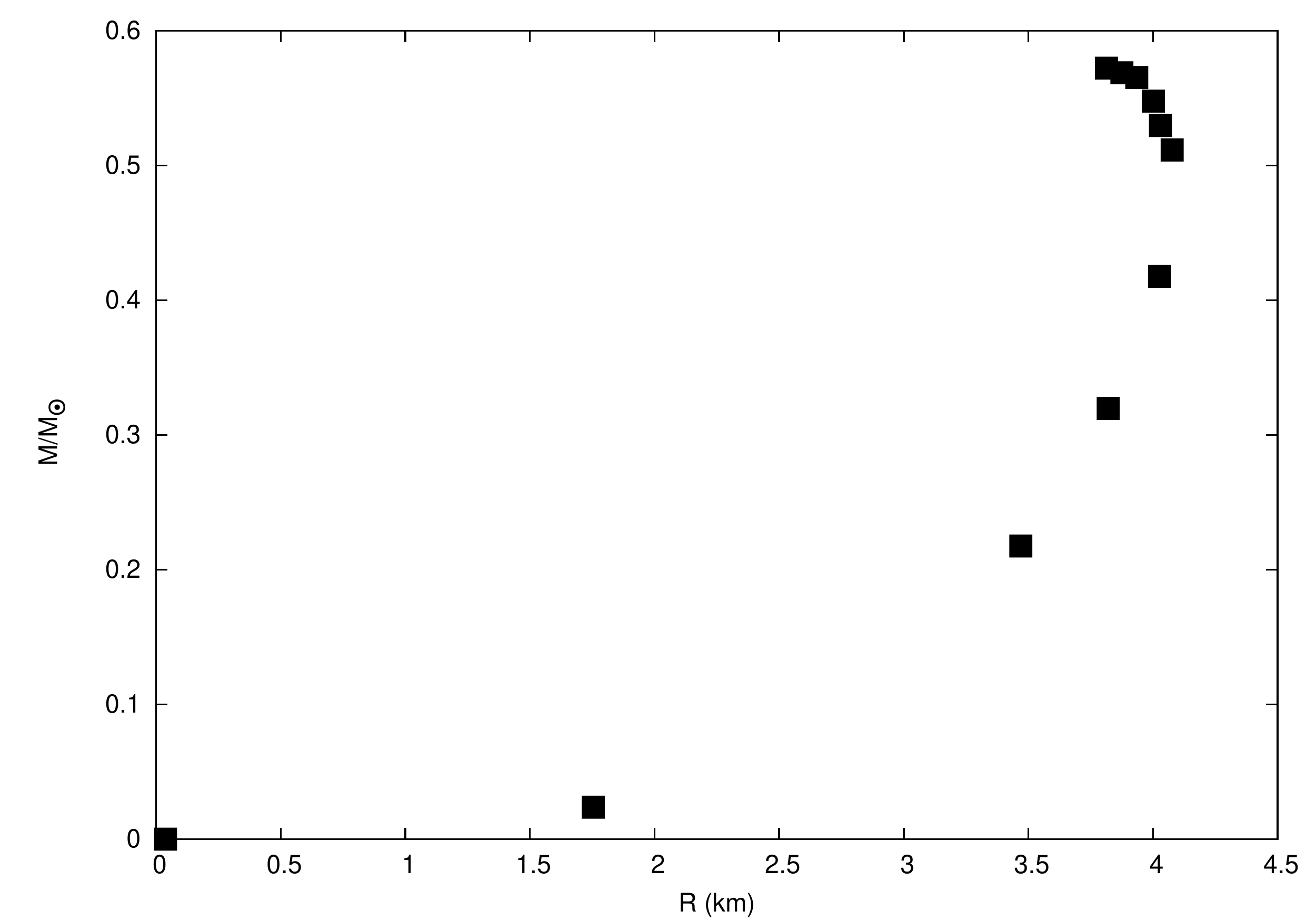}}
\caption{Neutron star mass in solar mass units as a function of the radius under the multilayered rational map approximation with shells showing non-equal baryon numbers. The dimensionless coupling constant is set to the physical value $\alpha = 7.3 \times 10^{-40}$ Plot obtained from data in Ref.~\refcite{Nelmes2011}.}\label{MvsR-Nelmes2011}
\end{figure}

The scheme observed in Ref.~\refcite{Nelmes2011} to solve the system can be summarized as follows: first, from an initial guess, a numerical configuration for the fields $g(y)$ and $q(y)$ is obtained by simulated annealing; and second, the Euler-Lagrange equations for $\nu(y)$ and $A(y)$ are solved through a forth-order Runge-Kutta method for given $N$ and $W$. This process is repeated by varying both $N$ and $W$ to find the minimal solution.

The corresponding results are shown in Fig. \ref{MvsR-Nelmes2011} by means of the mass-radius relation. They were achieved for a value of the coupling constant $\alpha = 7.3 \times 10^{-40}$ coming from the experimental pionic decay constant, $f_\pi = 186$ MeV, and $e = 4.84$ (following from fitting the $B=1$ Skyrmion in flat space to the proton mass). The maximum mass presents a baryon number $B = 2.84 \times 10^{56}$, an order of magnitude below the expected $B \sim 10^{57}$. The issue to reach high enough values of $B$ might be affected by the energies being overestimated, which is a known situation when dealing with Skyrmions in flat space under the rational map ansatz approximation.

For small baryon number, the radius grows with $B^{1/3}$, meaning a constant value of the average density. However, the latter goes up to 25 times when $B$ is close to its critical value. Indeed, the width of the shells increases with the radius as one should expect due to a lower compression from the gravitational interaction from the outer ones.

On the other hand, it is interesting to note that besides radial compression, there is lateral compression too. Although one could expect them to be equal, close to the center the star is more laterally compressed than radially, probably due to the multilayer structure of the ansatz. Obviously, the baryon density is larger at the center while decreasing along the radius of the star.

Concerning neutron stars' radius, the values achieved through this Skyrmionic approach are comparable to but lower than the typical 10 km radius of actual neutron stars. Furthermore, when close to the critical baryon number, the neutron star radius starts to decrease as an indicative signal of the dominance of the gravitational forces. Nevertheless, the values of the neutron star masses are still far from the big neutron star masses (up to 2.3 $M_\odot$) found in nature. All this discussion is manifest from Fig. \ref{MvsR-Nelmes2011}.

A final comment concerns the inclusion of the pion mass. Nelmes and Piette found in Ref.~\refcite{Nelmes2011} that, qualitatively, everything remains quite the same with a similar behavoir of the fields. However, the critical value of $B$ before solutions cease to exist slightly decreases from $B = 2.84 \times 10^{56}$ to $B = 2.72 \times 10^{56}$ with a reduction of the average shell width and a larger value of the energy per baryon.

Having a look at the mass {\it vs.} radius diagram, it is clear there is room for improvement in our description of neutron stars, specially, if we bear in mind that the results are based on the multilayered rational map ansatz -- a good approximation, but an approximation after all. In fact, this can be pursued by taking into account that, for large baryon number, the minimal configuration is better described by a cubic lattice of $B = 4$ Skyrmions. 

\subsubsection{Skyrmionic crystal}

Following this journey through a description of neutron stars based on the standard Skyrme model, our last stop corresponds to the Skyrmionic crystal. It is based on the fact that at large $B$, the configuration minimizing the static energy functional is given by a cubic lattice of $B = 4$ Skyrmions, {\it i.e.}, a Skyrmionic lattice of $\alpha$-particles. \cite{Castillejo1989} It is worth mentioning that previous attempts involving Skyrmion crystals have been carried out. For instance, T. Walhout considered a lattice of $B=1$ solitons either in a simple \cite{Walhout1988} or a face-centered \cite{Walhout1990} cubic lattice. After obtaining the corresponding equation of state, the Tolman-Oppenheimer-Volkoff (TOV) system \cite{Tolman1939,Oppenheimer1939} was solved giving high values of the neutron star masses: up to $3.63 M_\odot$ with a radius of 19.0 km in the simple cubic arrangement and $M = 2.57 M_\odot$ with $R = 15.2$ km for the face-centered cubic one.

Back to the $B=4$ lattice, we will review the mechanism followed by Nelmes and Piette in Ref.~\refcite{Nelmes2012} and their main results. Important to note is that instead of the usual TOV equations, we will consider the generalization described in Ref.~\refcite{Bowers1974} to account for anisotropic matter. As already mentioned, the lowest energy per baryon corresponds to a cubic lattice of face-centered Skyrmions with $B=4$, that is to say, $\alpha$-particles. Since in the absence of quantization we cannot distinguish between neutrons and protons, this construction in the neutron star framework will be equivalent to a fundamental cell made of four neutrons kept together.

To allow for anisotropy, we let the crystal deviate from a fundamental cubic cell of side $a$ to a rectangular one, introducing $p = r - 1/r$ as the corresponding measure of the deviation. Here, $r^3$ is the aspect ratio between the $x$ and $y$ directions with respect to $z$, {\it i.e.}, $x/z = r^3$ with $x=y$. Under these considerations, Castillejo et al. in Ref.~\refcite{Castillejo1989} found an expression for the energy of a single Skyrmion with massless pions in terms of $p$ and its size $L=n^{-1/3}$, with $n$ being the Skyrmion number density:
\begin{equation}\label{E_lattice}
E (L,p) = E_{p=0} (L) + E_0 [\, \alpha (L) \, p^2 + \beta (L) \, p^3 + \gamma (L) \, p^4 + \delta (L) \, p^5 + \dots],
\end{equation}

\noindent with the defined quantities
\begin{equation}
E_{p=0} (L) = E_0 \left[ 0.474 \left( \frac{L}{L_0} + \frac{L_0}{L} \right) + 0.0515 \right],
\end{equation}
\begin{equation}
\alpha (L) = 0.649 - 0.487 \frac{L}{L_0} + 0.089 \frac{L_0}{L},
\end{equation}
\begin{equation}
\beta (L) = 0.300 + 0.006 \frac{L}{L_0} - 0.119 \frac{L_0}{L},
\end{equation}
\begin{equation}
\gamma (L) = -1.64 + 0.78 \frac{L}{L_0} + 0.71 \frac{L_0}{L},
\end{equation}
\begin{equation}
\delta (L) = 0.53 - 0.55 \frac{L}{L_0},
\end{equation}

\noindent where $E_0 = 727.4$ MeV and $L_0 = 1.666 \times 10^{-15}$ m. As expected, the minimum of the energy for arbitrary $L$ corresponds to $p=0$, that is to say, to the face-centered cubic crystal, with $L = L_0$ the global minimum. 

In addition, if $\lambda_r$ and $\lambda_t$ are the lengths of the Skyrmion in the radial and tangential directions of the neutron star, respectively, the quantities $L$ and $p$ read
\begin{equation}
L = (\lambda_r \lambda_t \lambda_t)^{1/3}, \qquad p =\left( \frac{\lambda_t}{\lambda_r} \right)^{1/3} - \left( \frac{\lambda_r}{\lambda_t} \right)^{1/3}.
\end{equation}

Bearing in mind that we want our configurations to be spherically symmetric but also anisotropic in general, the energy-momentum tensor we are interested in should take the form
\begin{equation}
T^\mu_\nu = {\rm diag}(\rho(r), p_r(r), p_t(r), p_t(r)),
\end{equation}

\noindent with $\rho(r)$ the energy density, and $p_r(r)$ and $p_t(r)$ the radial and tangential pressure, respectively.

On the other hand, besides the EoS, we need a generalized TOV equation accounting for this anisotropy.\cite{Bowers1974} Hence, we consider the usual spherically symmetric metric but written in a slightly different form to match the one used by Nelmes and Piette,
\begin{equation}
ds^2 = e^{\nu(r)} dt^2 - e^{\lambda(r)} dr^2 - r^2( d\theta^2 + \sin^2 \theta \, d\phi^2).
\end{equation}

\noindent Together with the energy-momentum tensor, we arrive at Einstein's equations:
\begin{equation}\label{EinsteinLattice1}
e^{-\lambda} \left(\frac{\lambda'}{r}-\frac{1}{r^2} \right) + \frac{1}{r^2} = 8 \pi \rho,
\end{equation}
\begin{equation}\label{EinsteinLattice2}
e^{-\lambda} \left(\frac{\nu'}{r}+\frac{1}{r^2} \right) - \frac{1}{r^2} = 8 \pi p_r,
\end{equation}
\begin{equation}\label{EinsteinLattice3}
e^{-\lambda} \left(\frac{1}{2} \nu'' - \frac{1}{4} \lambda' \nu' + \frac{1}{4} \nu'^2 + \frac{\nu' - \lambda'}{2 r} \right) = 8 \pi p_t.
\end{equation}

Defining the gravitational mass of the neutron star up to a radius $r$
\begin{equation}
m(r) =\int_0^r 4 \pi r^2 \rho \, dr,
\end{equation}

\noindent and integrating the first Einstein equation, we are led to
\begin{equation}
e^{-\lambda} = 1 - \frac{2m}{r}.
\end{equation}

\noindent Next, we can plug this expression in Eq. (\ref{EinsteinLattice2}) getting
\begin{equation}\label{nuEq}
\frac{1}{2} \nu' = \frac{m + 4 \pi r^3 p_r}{r(r - 2m)}.
\end{equation}

\noindent And by differentiating Eq. (\ref{EinsteinLattice2}) together with Eq. (\ref{EinsteinLattice3}), we obtain
\begin{equation}\label{prEq}
\frac{dp_r}{dr} = - (\rho + p_r) \frac{\nu'}{2} + \frac{2}{r} (p_t - p_r).
\end{equation}

\noindent Finally, combining Eq. (\ref{nuEq}) with Eq. (\ref{prEq}), we arrive at
\begin{equation}\label{AnisotrpicTOV}
\frac{dp_r}{dr} = - (\rho + p_r) \frac{m + 4 \pi r^3 p_r}{r(r - 2m)} + \frac{2}{r} (p_t - p_r),
\end{equation}

\noindent which is the generalized TOV equation.

To solve the TOV system, two equations of state have to be provided, namely, $p_r= p_r (\rho)$ and $p_t = p_t(\rho)$. At zero temperature, they can be easily obtained from the energy per Skyrmion given in Eq. (\ref{E_lattice}): \cite{Nelmes2012}
\begin{equation}\label{LatticeEoS}
p_r = - \frac{1}{\lambda_t^2} \frac{\partial E}{\partial \lambda_r}, \qquad p_t = - \frac{1}{\lambda_r} \frac{\partial E}{\partial \lambda_t^2}.
\end{equation}

\noindent In addition, boundary conditions are needed. Due to the fact that at the center there is no matter enclosed, $m(r = 0) = 0$. Furthermore, $p_r(r = 0)$  and $dp_r/dr |_{r=0}$ have to be finite [see Eq. (\ref{nuEq}) and (\ref{prEq})], which imply that $\nu'(r=0) = 0$ and $p_t(r =0) = p_r(r=0)$, respectively.

The radius $R$ of the neutron star is defined by the condition $p_r(R) = 0$, since the external pressure on the surface is expected to be negligible. Nevertheless, this is not the case for $p_t$. The only required condition is $p_r, \, p_t \geq 0, \, \forall r \leq R$.

Thus, for a fixed $B$, the minimal configuration will be that with a minimum value of the gravitational mass
\begin{equation}
M_G = m(R) = \int_0^R 4 \pi r^2 \rho \, dr,
\end{equation}

\noindent and 
\begin{equation}
\rho = \frac{E}{\lambda_r \lambda_t^2 c^2}.
\end{equation}

\noindent Then, the procedure to find solutions consists in minimizing $M_G$ for $\lambda_r(r)$ and $\lambda_t(r)$. The method using a simulated annealing algorithm is described in Ref.~\refcite{Nelmes2012}. The main idea is that if one has a profile $\lambda_t(r)$, it is possible to infer from it the corresponding $\lambda_r$ as a function of $r$ and integrate $m(r)$ up to the radius $R$ to obtain the neutron star mass $M_G$. In this way, the problem is reduced to minimize $M_G$ with respect $\lambda_t(r)$ by using simulated annealing. At the same time, we need to take care of the baryon number, and at each step, $\lambda_t$ must be rescaled to restore the desired value of $B$, repeating the integration of $M_G$ until no rescaling is needed. For this purpose, one should consider that the baryon number is given by
\begin{equation}
B = \int_0^R \frac{4 \pi r^2 n(r)}{\left(1 - \frac{2 G m}{c^2 r} \right)^{1/2}} \,  dr,
\end{equation}

\noindent with
\begin{equation}
n(r) = \frac{1}{\lambda_r(r) \lambda_t^2(r)}
\end{equation}

\begin{figure}[th]
\centerline{\includegraphics[width=12cm]{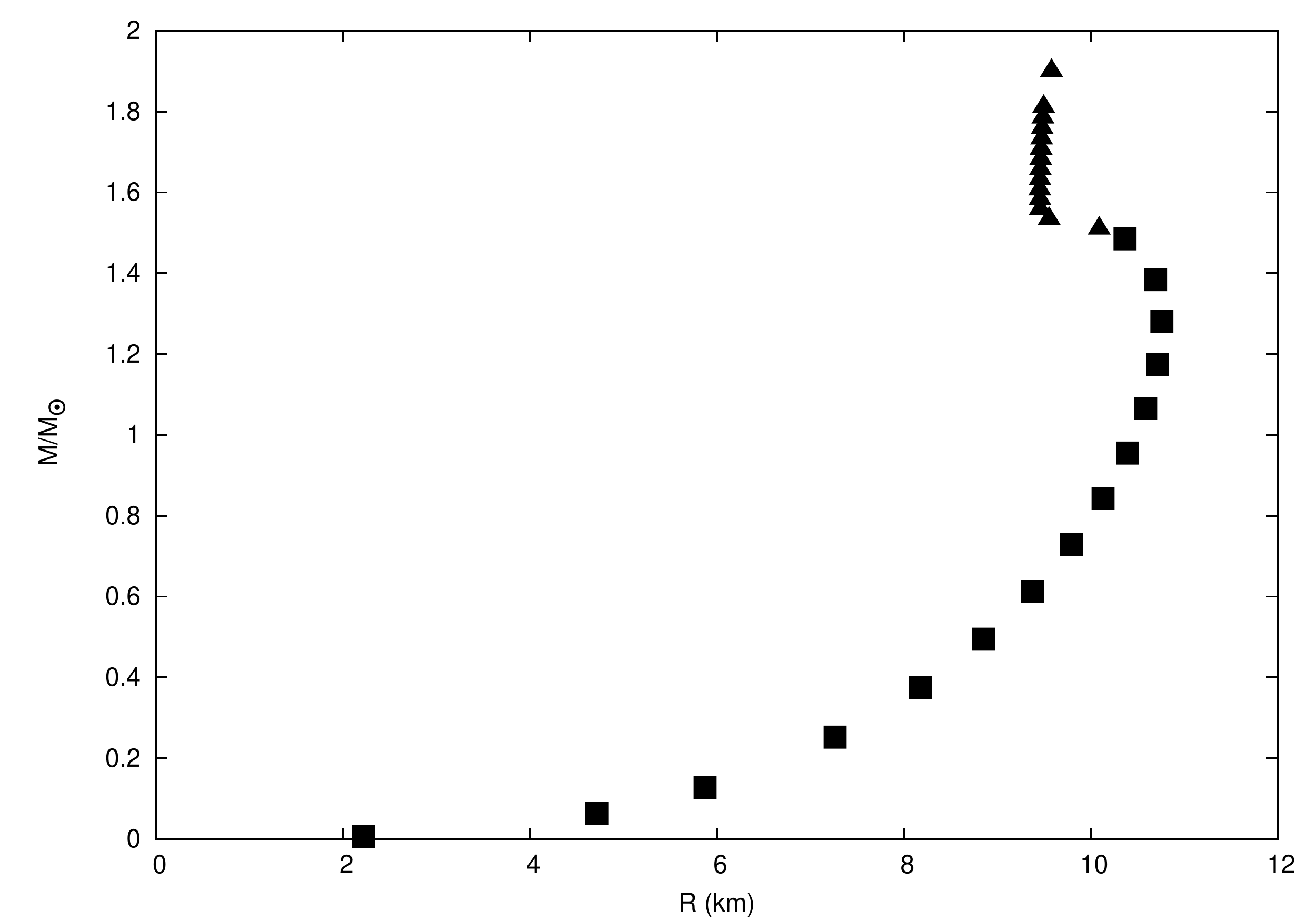}}
\caption{Neutron star mass in solar mass units as a function of the radius for the $B=4$ Skyrmionic crystal. The square (triangle) points correspond to isotropic (anisotropic) neutron stars. Plot obtained from data in Ref.~\refcite{Nelmes2012}.}\label{MvsR-Nelmes2012}
\end{figure}

The main result concerning the description of neutron stars by means of Skyrmionic crystals can be summarized as follows: For baryon number up to $2.61 \times 10^{57}$ (corresponding to a gravitational mass of $1.49 M_\odot$), we find isotropic neutron stars. From this value and up to $B = 3.25 \times 10^{57}$ ($M_G = 1.81 M_\odot$), solutions still exist but made of anisotropically deformed matter (see Fig. \ref{MvsR-Nelmes2012}). As usual, the configurations are stable when the energy per baryon decreases when increasing the baryon number and are unstable otherwise.

From the mass-radius diagram in Fig. \ref{MvsR-Nelmes2012}, we see that those solutions with masses bigger than the solar mass present radii of about 10 km, while for isotropically deformed neutron stars above $1.28 M_\odot$, the radius starts to decrease. In the anisotropic case, there is an important drop of the radius which keeps an almost constant value of about 9.5 km.

As one could expect, Skyrmions are more compressed towards the center than at the surface of the neutron star, which is translated into the energy density: highest at the center and decreasing along the radial direction, due to the effect of the outer matter compressing the inner parts.

The inner structure of the neutron star may be analized by focusing on $\lambda_r(r)$ and $\lambda_t(r)$. As supported by Fig. 5 in Ref.~\refcite{Nelmes2012}, at the center, $\lambda_r = \lambda_t$, a behavior holding along the configuration for the isotropic solution. In the anisotropic case, the anisotropy increases in the radial direction, taking its maximal value on the surface. It is here where solitons show a larger tangential deformation. 

We can also study $\lambda_t$ and $\lambda_r$ both at the center and on the star surface as a function of the gravitational mass (for a graphical description, the reader can look for Fig. 7 in Ref.~\refcite{Nelmes2012}). When the neutron star is isotropic, $\lambda_r(0) = \lambda_t(0)$ monotonously decreases while the mass increases, showing that the density at the center also grows with the mass. On the other hand, $\lambda_{t,r}(R)$ remain constant with the same value as in the absence of gravity since there is no matter compressing them. However, after the phase transition there is an abrupt decrease to a minimum value of $\lambda_{t,r}$ at the center with $\lambda_{t,r}(R)$ keeping almost the same value. After it, $\lambda_{t,r}(0)$ start to increase with both $\lambda_t(R)$ and $\lambda_r(R)$ decreasing, although with $\lambda_t(R)$ in a quicker way, showing that for these neutron stars, the compression along the tangential direction is much bigger than in the radial one.

Finally, we can add the pion mass contribution already shown in the Lagrangian of the standard Skyrme model in Eq. (\ref{StandardLag}). In the cubic lattice of $B=4$ Skyrmions it is translated into a term proportional to the soliton volume,
\begin{equation}
E_\pi= \frac{1}{4} m_\pi^2 f_\pi^2 L^3.
\end{equation}

\noindent The effect of massive pions consists in slightly reducing the maximum mass neutron stars can achieve: $M_G = 1.47 M_\odot$ for isotropic and $M_G = 1.88 M_\odot$ for anisotropic solutions.

It is not surprising we get a better description of neutron stars by using the Skyrmionic crystal configuration. This is the true minimizer of the Skyrme model at high baryon number and there is no bigger object in nature (in terms of $B$) than neutron stars. However, besides having problems to reach high values of their masses, the lattice arrangement is not a good description for their inner structure, which is thought to be of the superfluid type, and constitutes the main and large part of the star (the typical crust width presents a value of 1 or 2 km to be compared to a radius of the order of 12 km \cite{Lattimer2004}). 

It is in helping us achieve a better description on both the maximum star mass and its structure where the other Skyrme theory we have presented in Section \ref{BPS-Model}, the BPS model, will have a prominent role.


\subsection{Neutron stars in the BPS Skyrme model}

The coupling of the BPS Syrme model to gravity is analogous to that given by Eq. (\ref{S-gravity}). In this case, the matter part of the action for a general metric $g_{\mu \nu}$ reads \cite{Adam2015a,Adam2015b}
\begin{equation}
S_{\rm BPS} = \int |g|^\frac{1}{2} \left( - \lambda^2 \pi^4 |g|^{-1} g_{\mu \nu} \mathcal B^\mu \mathcal B^\nu - \mu^2 \mathcal U \right) d^4 x,
\end{equation}

\noindent where $\mathcal U$ corresponds to an arbitrary potential. The energy-momentum tensor in curved space-time still corresponds to that of a perfect fluid, namely,
\begin{equation}
T^{\mu \nu} = - 2 |g|^{-\frac{1}{2}} \frac{\delta}{\delta g_{\mu \nu}} S_{\rm BPS} = (p + \rho) u^\mu u^\nu - p \, g^{\mu \nu},
\end{equation}

\noindent and in this general framework, the 4-velocity is given by
\begin{equation}
u^\mu = \frac{\mathcal B^\mu}{\sqrt{g_{\rho \sigma} \mathcal B^\rho \mathcal B^\sigma}},
\end{equation}

\noindent whereas for the energy density and pressure we have
\begin{equation}
\rho = \lambda^2 \pi^4 |g|^{-1} g_{\mu \nu} \mathcal B^\mu \mathcal B^\nu + \mu^2 \mathcal U,
\end{equation}
\begin{equation}
p = \lambda^2 \pi^4 |g|^{-1} g_{\mu \nu} \mathcal B^\mu \mathcal B^\nu - \mu^2 \mathcal U.
\end{equation}

\noindent Note that the baryon density, $\mathcal B^\mu$, is defined as in Eq. (\ref{Bmu}).

To simplify the study, it is common practice in the BPS Skyrme model to consider the parametrization of the $U$ field ($\vec \tau$ denotes the triple of Pauli matrices),
\begin{equation}
U = e^{i \xi \vec n \cdot \vec \tau} = \cos \xi + i \sin \xi \vec n \cdot \vec \tau, \qquad \vec n^2 =1,
\end{equation}

\noindent together with an axially symmetric ansatz, which in spherical coordinates reads
\begin{equation}
\xi = \xi(r), \qquad \vec n = (\sin \theta \cos (B\phi), \sin \theta \sin(B\phi), \cos \theta),
\end{equation}

\noindent with $B$ the total baryon number. This choice is usually taken by its simplicity. It is possible because the static energy functional of the BPS model has the volume preserving diffeomorphisms as a symmetry, so the specific shape of the configurations does not change the energy and thus, it is not important for its minimization. In addition, it is also the best ansatz to describe neutron stars due to the effect of gravity, which favors the spherical symmetry.

Indeed, this leads us to consider again a spherically symmetric metric. To match the work in Refs.~\refcite{Adam2015a,Adam2015b}, we choose the implementation
\begin{equation}
ds^2 = \A(r) dt^2 - \B(r) dr^2 -r^2 (d\theta^2 + \sin^2 \theta d\phi^2),
\end{equation}

\noindent where the bold $\A$ and $\B$ are not vectors but the metric fields (this notation is used to avoid confusion with the mass or baryon number). This allows us to calculate the Ricci scalar
\begin{equation}
R = \frac{\A''}{\A \B} - \frac{1}{2} \frac{\A' \B'}{\A \B^2} - \frac{1}{2} \frac{(\A')^2}{\A^2 \B} + \frac{2}{r} \frac{\A'}{\A \B} - \frac{2}{r^2} \left(1 -\left(\frac{r}{\B} \right)' \right),
\end{equation}

\noindent finally arriving at Einstein's equations
\begin{equation}\label{BPS-Einstein1}
\frac{1}{r} \frac{\B'}{\B} = -\frac{1}{r^2} (\B -1) + \frac{\kappa^2}{2} \B \rho,
\end{equation}
\begin{equation}\label{BPS-Einstein2}
r (\B p)' = \frac{1}{2} (1 - \B) \B (\rho + 3 p) + \frac{\kappa^2}{2} \mu^2 r^2 \B^2 \mathcal U(h) p,
\end{equation}
\begin{equation}\label{BPS-Einstein3}
\frac{\A'}{\A} = \frac{1}{r} (\B - 1) + \frac{\kappa^2}{2} r \B p,
\end{equation}

\noindent where $\kappa^2 = 16 \pi G$, being $G$ the Newton gravitational constant. In addition, a new field $h(r) = \frac{1}{2}(1 - \cos(\xi(r)))$ has been defined, with $\rho$ and $p$ taking the form
\begin{equation}\label{rhoFULL}
\rho = \frac{4 B^2 \lambda^2}{\B \, r^4} h (1-h) h'^2 + \mu^2 \mathcal U(h),
\end{equation}
\begin{equation}\label{pFULL}
p = \rho - 2 \mu^2 \mathcal U(h) = \frac{4 B^2 \lambda^2}{\B \, r^4} h (1-h) h'^2 - \mu^2 \mathcal U(h).
\end{equation}

\noindent All in all, we have a system of three differential equations where the field $\A$ is decoupled. Then, the first two can be solved to get $\B$ and $h$, with the third one giving $\A$ in terms of the former fields.

To find the solutions, we will perform a numerical calculation, concretely, we will use a shooting from the center with a fourth-order Runge-Kutta method. To proceed with this scheme, we first need the value of the two parameters in this BPS theory, {\it i.e.}, $\lambda$ and $\mu$. In this case, since we are dealing with neutron stars, it seems a better choice \cite{Adam2015b} to fit the parameters to both the binding energy per nucleon of infinite nuclear matter and the nuclear saturation density:
\begin{equation}
E_b = 16.3 \; {\rm MeV}, \qquad n_0 = 0.153 \; {\rm fm}^{-3}
\end{equation}

Due to the fact that $\mu^2$ is the coupling corresponding to the potential, the fit will depend on this particular choice. In the following, we will focus on three different options, namely, the step function potential, the pion mass potential and the pion mass potential squared. By using Eq. (\ref{BPS_E}) and (\ref{BPS_EoS}) given in section \ref{BPS-Model} for the energy and volume, we obtain
\begin{eqnarray}
\Theta(h)&:& \quad E = 2 \pi^2 B \lambda \mu, \qquad V = \pi^2 B \frac{\lambda}{\mu},    \\
\mathcal U_\pi = 2 h&:&  \quad E = \frac{64 \sqrt 2 \pi}{15} B \lambda \mu, \qquad V = \frac{8}{3} \sqrt 2 \pi B \frac{\lambda}{\mu},  \\
\mathcal U_\pi^2 = 4h^2&:& \quad E = 2 \pi^2 B \lambda \mu, \qquad V = 2 \pi^2 B \frac{\lambda}{\mu},
\end{eqnarray}

\noindent which lead to the fitted values displayed in Table \ref{Fit}. 

\begin{table}[pt] \label{Fit}
\caption{Parameter values from the fit to $E_b$ and $n_0$.}
\begin{center}
\begin{tabular}{@{}ccc@{}} \hline
Potential & $\lambda^2$ (MeV ${\rm fm^3}$) & $\mu^2$ (MeV ${\rm fm^{-3}}$) \\ \hline
$\Theta (h)$ & 30.99 & 70.61 \\
$\mathcal U_\pi = 2 h$ & 26.88 & 88.26 \\
$\mathcal U_\pi^2 = 4 h^2$ & 15.493 & 141.22
\\ \hline
\end{tabular}\label{Fit}
\end{center}
\end{table}

As already mentioned, the solvability of the BPS model will allow us to bring together two different approaches for a solitonic description of neutron stars. On the one hand, we are able to consider the full field theory defined by Einstein's equations (\ref{BPS-Einstein1}), (\ref{BPS-Einstein2}) and (\ref{BPS-Einstein3}) with the back-reaction of gravity included. On the other hand, we may also use the mean equation of state given in Eq. (\ref{MF_rho}) together with the corresponding TOV equations to study the commonly assumed mean field limit. One should note that for the step functional potential, both approaches are exactly the same.

In the first case, we just numerically integrate the three Einstein equations with boundary conditions at the center,
\begin{equation}
h(r = 0) = 1, \qquad \B(r =0) =1.
\end{equation}

\noindent These values come from the facts that the profile function at the soliton center takes the value $\pi$ and that there is no matter enclosed, respectively. From the expansion of the $h$ field at the origin, $h \sim 1 - \frac{1}{2} h_2 r^2$, one may see there is only one free parameter left, $h_2$, which, by Eq. (\ref{rhoFULL}), is equivalent to the energy density at the center of the star:
\begin{equation}
\rho_0 \equiv \rho (r = 0) = B^2 \lambda^2 h_2^3 + \mu^2 \mathcal U (h =1).
\end{equation}

\noindent On the other hand, its radius $R$ is defined as the point where the matter field $h$ vanishes, {\it i.e.}, $h(R) = 0$, which at the same time also implies that the pressure is equal to zero [see Eq. (\ref{pFULL}) together with $\mathcal U(h=0) = 0$].

There is also an extra condition coming from Eq. (\ref{BPS-Einstein2}). For a non-singular metric $\B$ at the neutron star surface, it implies that not only the pressure but also its derivative has to vanish, {\it i.e.}, $p'(R) = 0$. Hence, this is translated into
\begin{equation}
\frac{4 B^2 \lambda^2}{\B(R) R^4} h_r^2 (R) - \mu^2 \mathcal U_h(0) = 0,
\end{equation}

\noindent for the derivative of the matter field $h$ at $r = R$.

In addition, for $r \geq R$, the configuration is joined to the vacuum solution
\begin{equation}
h(r) = 0, \qquad \B = \left(1 - \frac{2 G M}{r}\right)^{-1},
\end{equation}

\noindent where $M$ is the total mass.

Thus, for a fixed baryon number, one varies the energy density at the center to find solutions satisfying these constraints. Regarding the baryon number, the situation is as follows: for small values there is only one solution available; however, by increasing $B$, two branches of configurations appear until we arrive at a maximum value $B_{\rm max}$ from where solutions cease to exist. This is not a new behavior. Indeed, it is also the case for the well-known TOV system of a free relativistic fermi gas \cite{Weinberg1972}, and in a similar way, the branch corresponding to higher values of the central energy density is considered unstable.

Focusing  on the mean field limit, the TOV system is again given by Eq. (\ref{BPS-Einstein1}) and (\ref{BPS-Einstein2}) with the consideration that $2 \mu^2 \mathcal U(h) = \rho -p$. Nevertheless, the energy density and the pressure have to be thought now as independent variables instead of functions of the matter field $h$, so we have a system of three equations and three variables: $\bar \rho$, $p$ and $\B$, where the change of notation accounts for the mean field nature of the system. The initial condition on the metric field is still the same, $\B(r=0)=1$, and there is one free parameter which will allow us to get different solutions: the energy density at the center, $\bar \rho(0) = \bar \rho_0$.

Therefore, for different values of $\bar \rho_0$, we can perform a shooting from the center to obtain the desired solutions. In this process, we find stable solutions as long as increasing the value of the energy density at the centre gives us neutron stars with higher values of their mass; otherwise, the solution is unstable. The neutron star mass is given by
\begin{equation}
M = 4 \pi \int_0^R dr \, r^2 \bar \rho (r),
\end{equation}

\noindent while its radius $R$ is defined now by the condition $p(R) = 0$. In addition, it is important to note that the baryon number is not a free parameter anymore as in the full calculation. Instead, it is obtained {\it a posteriori} by the integral
\begin{equation}
B = 4 \pi \int_0^R dr \, r^2 \sqrt{\B} \bar \rho_B,
\end{equation}

\noindent with $\bar \rho_B$ the mean field baryon density, which from the expression for the volume in Eq. (\ref{BPS_EoS}) reads
\begin{equation}
\bar \rho_B = \frac{B}{V} = \frac{\mu}{\pi^2 \lambda \left \langle \frac{1}{\sqrt{\mathcal U + P/\mu^2}} \right \rangle},
\end{equation}

\noindent as an average over the target space.

\begin{figure}[t]
\centerline{\includegraphics[width=12cm]{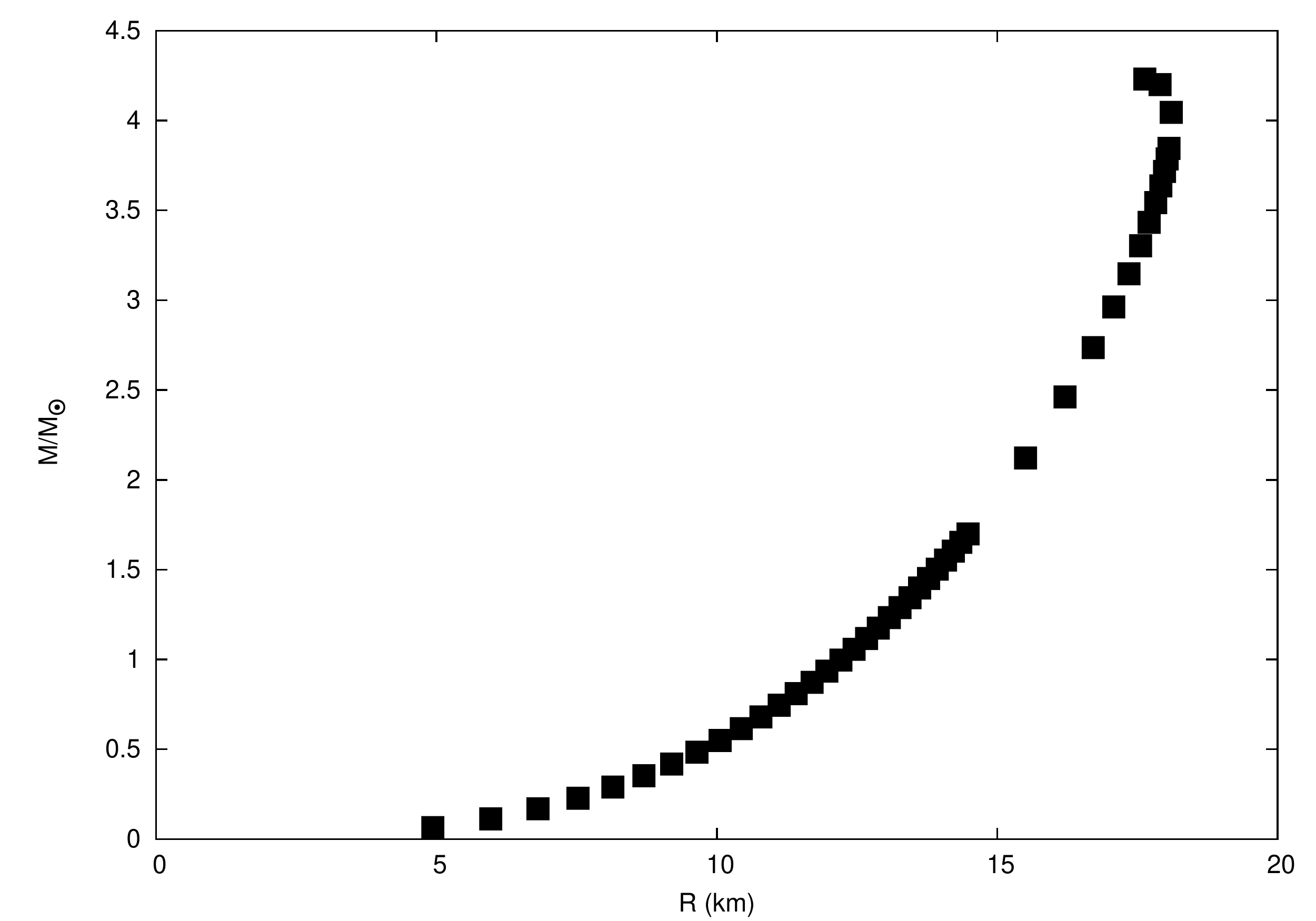}}
\caption{Neutron star mass in solar mass units as a function of the radius for the step function potential in the BPS model.}\label{MR-step}
\end{figure}

\begin{figure}[t]
\centerline{\includegraphics[width=12cm]{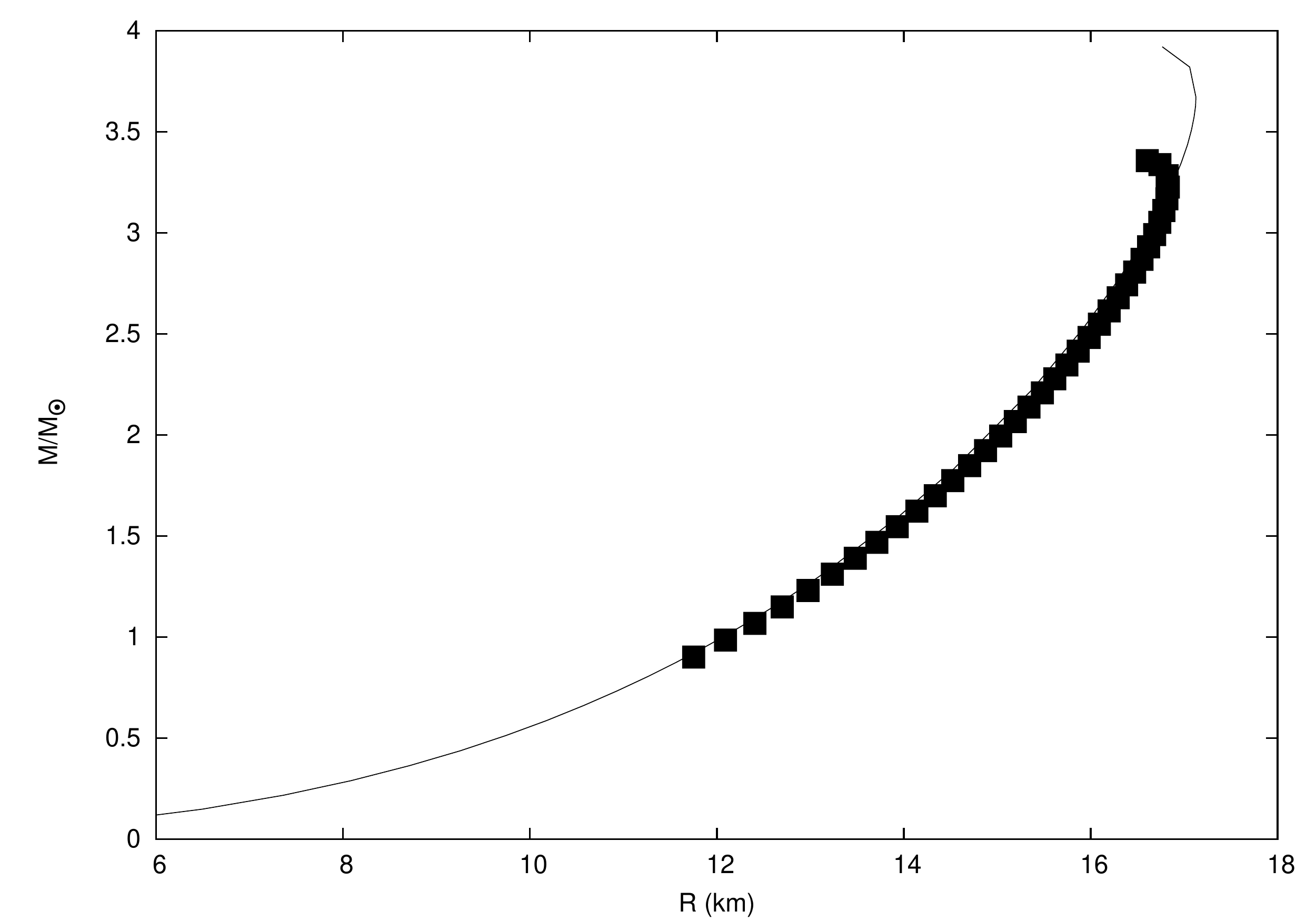}}
\caption{Neutron star mass in solar mass units as a function of the radius for the pion mass potential in the BPS model. The squared points correspond to the full calculation and the line to the mean field approach.}\label{MR-h}
\end{figure}

\begin{figure}[t]
\centerline{\includegraphics[width=12cm]{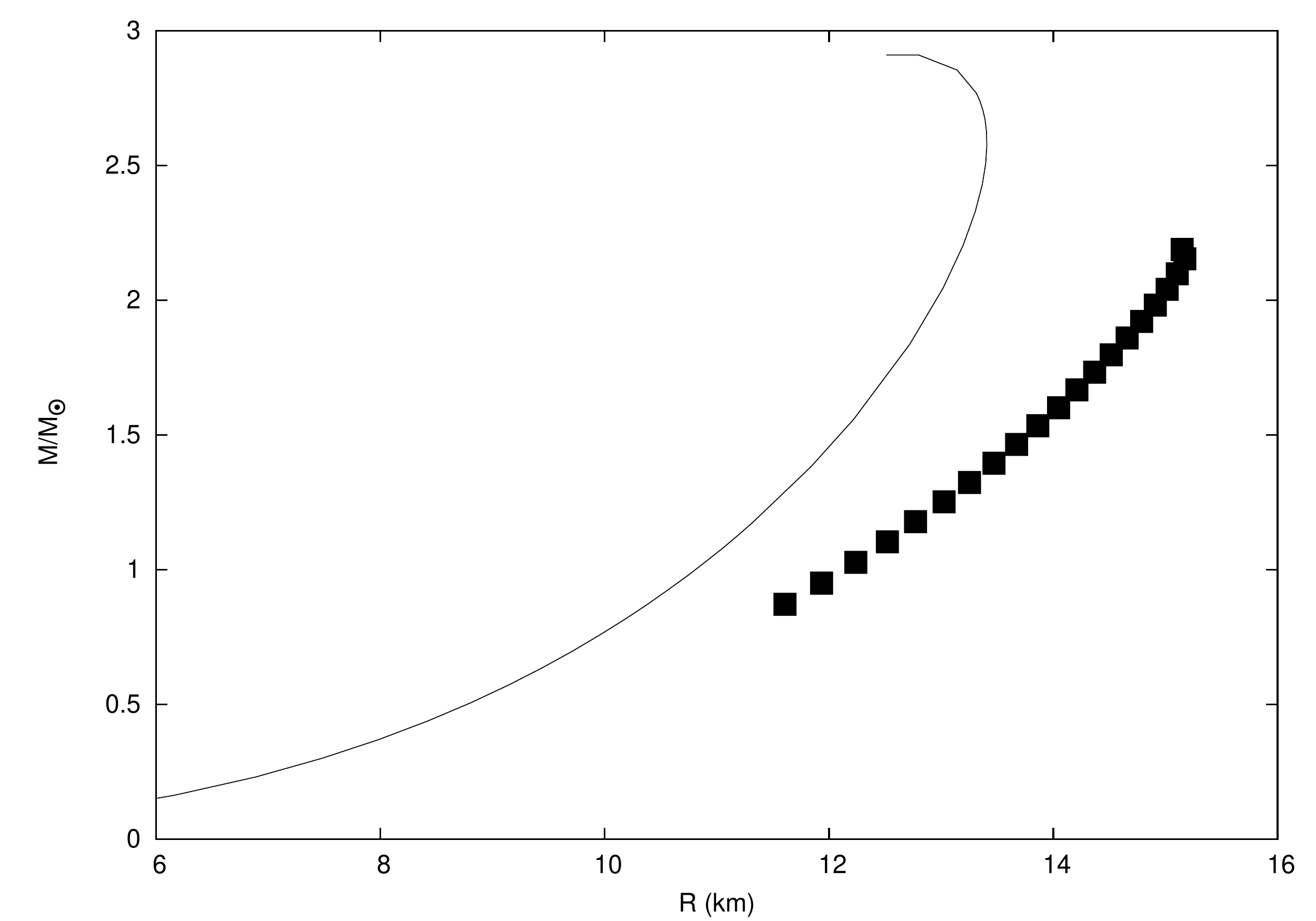}}
\caption{Neutron star mass in solar mass units as a function of the radius for the pion mass potential squared in the BPS model. The squared points correspond to the full calculation and the line to the mean field approach.}\label{MR-h2}
\end{figure}

We can now discuss the results obtained in Ref.~\refcite{Adam2015b} where the two approaches were confronted. In Figs. \ref{MR-step}, \ref{MR-h} and \ref{MR-h2} we present the mass-radius diagram for the three different potentials we have already introduced: the step function, the pion mass and the pion mass squared, respectively. For the step function potential shown in Fig. \ref{MR-step} both approaches agree since in this particular case the mean field approximation is equivalent to the full calculation. In this case, the mean field equation of state as given by Eq. (\ref{MF_rho}) is
\begin{equation}
\bar \rho = P + 2 \mu^2,
\end{equation}

\noindent providing a maximal neutron star mass of $M = 4.23 M_\odot$.

In the case of the pion mass potential, the mean field EoS corresponds to
\begin{equation}
\bar \rho = \frac{\mu^2}{5} \left(2 - 3 \frac{P}{\mu^2} + \frac{6}{1 + \frac{P}{\mu^2} \left(1- \frac{K\left[\frac{2}{2 + P/\mu^2} \right]}{E\left[\frac{2}{2 + P/\mu^2} \right]} \right)} \right),
\end{equation}

\noindent where $K$ and $E$ are the elliptic integrals of the first and second kind, respectively. This allows for neutron stars up to $3.79 M_\odot$ which can be compared with the value $M = 3.34 M_\odot$ coming from the full calculation. In Fig. \ref{MR-h}, it can be seen that, although the masses are slightly different, the mass-radius diagram is quite similar for both approaches. This situation does not appear when the potential is the pion mass squared.

In Fig. {\ref{MR-h2} one can observe that the two approaches differ considerably. This is due to the fact that the potential is more peaked around the center so the system deviates further from the mean field approximation, whose equation of state reads
\begin{equation}
\bar \rho = \mu^2 \left(\frac{P}{\mu^2} + \frac{5 \, _3F_2\left[\frac{1}{2},\frac{7}{4},\frac{9}{4};\frac{5}{2},3;-\frac{4\mu^2}{P}\right]}{2\, _3F_2\left[\frac{1}{2},\frac{3}{4},\frac{5}{4};\frac{3}{2},2;-\frac{4\mu^2}{P}\right]} \right),
\end{equation}

\noindent with $_pF_q[a_1,\ldots, a_p; b_1, \ldots, b_q; z]$ the generalization of the hypergeometric function. This is translated not only into a lower maximal mass for the exact solution, {\it i. e.}, $M = 2.15 M_\odot$ comparing to $M = 2.82 M_\odot$ of the mean field approach, but also into a larger neutron star radius.

One common property to all diagrams is that, in general, the star radius $R$ increases with the mass $M$ except very close to the maximal masses. This effect appears both in the full calculation and the mean field approach and is due to the stiffness of the equation of state. On the other hand, the masses are higher in the mean field system than in the exact calculation. Notwithstanding, we can say that the global properties of neutron stars are approximately the same. 

\begin{figure}[t]
\centerline{\includegraphics[width=12cm]{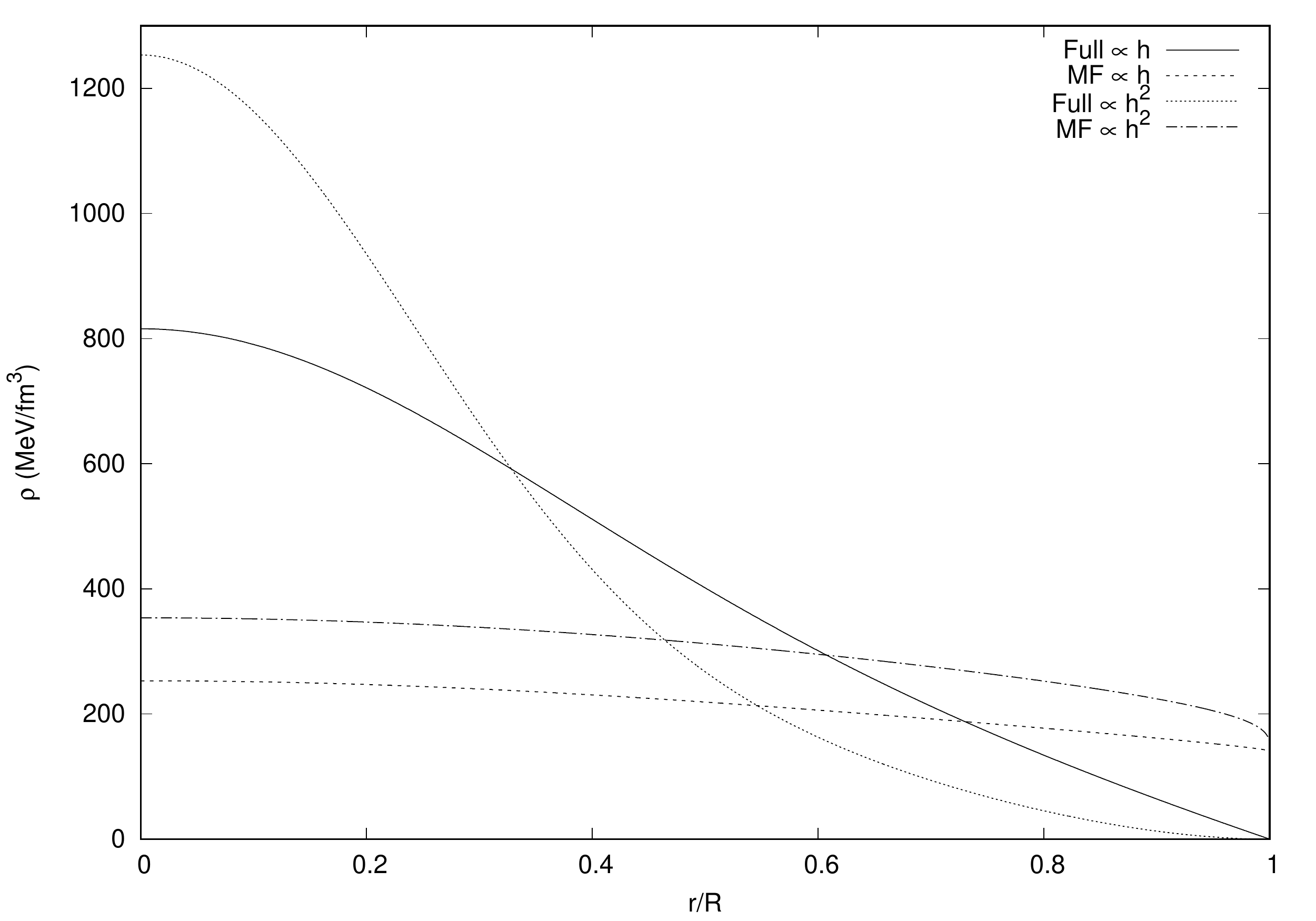}}
\caption{Energy density as a function of the radial coordiante $r$ normalised by the neutron star radius $R$. The solid and dashed lines correspond to the maximal mass for the standard pion mass potential. The dotted and dashed-dotted lines correspond to the maximal mass for the pion mass potential squared. }\label{EvsR}
\end{figure}

\begin{figure}[h!] \centering
\subfigure[]{\includegraphics[width=.7\textwidth]{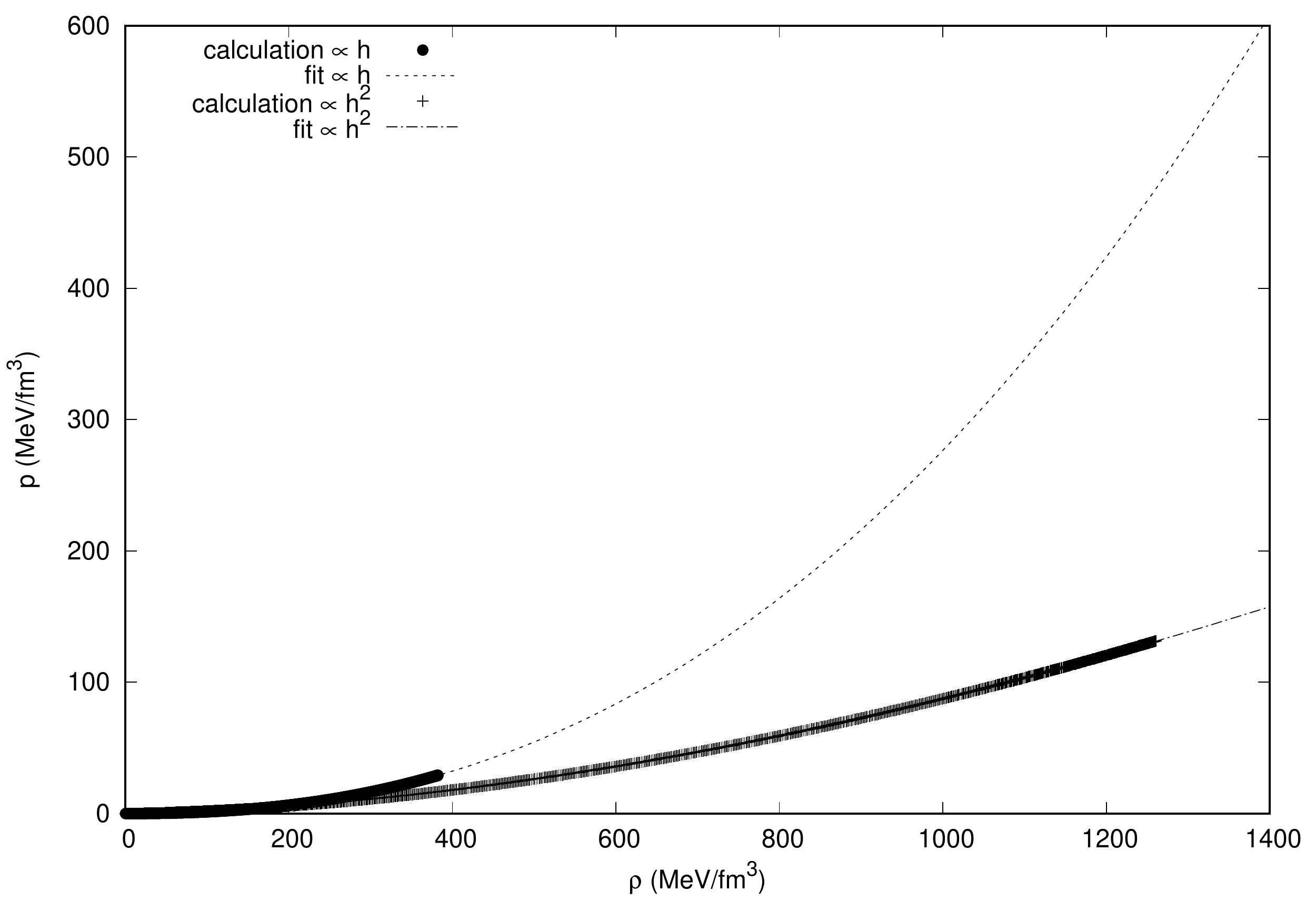}}
\subfigure[]{\includegraphics[width=.7\textwidth]{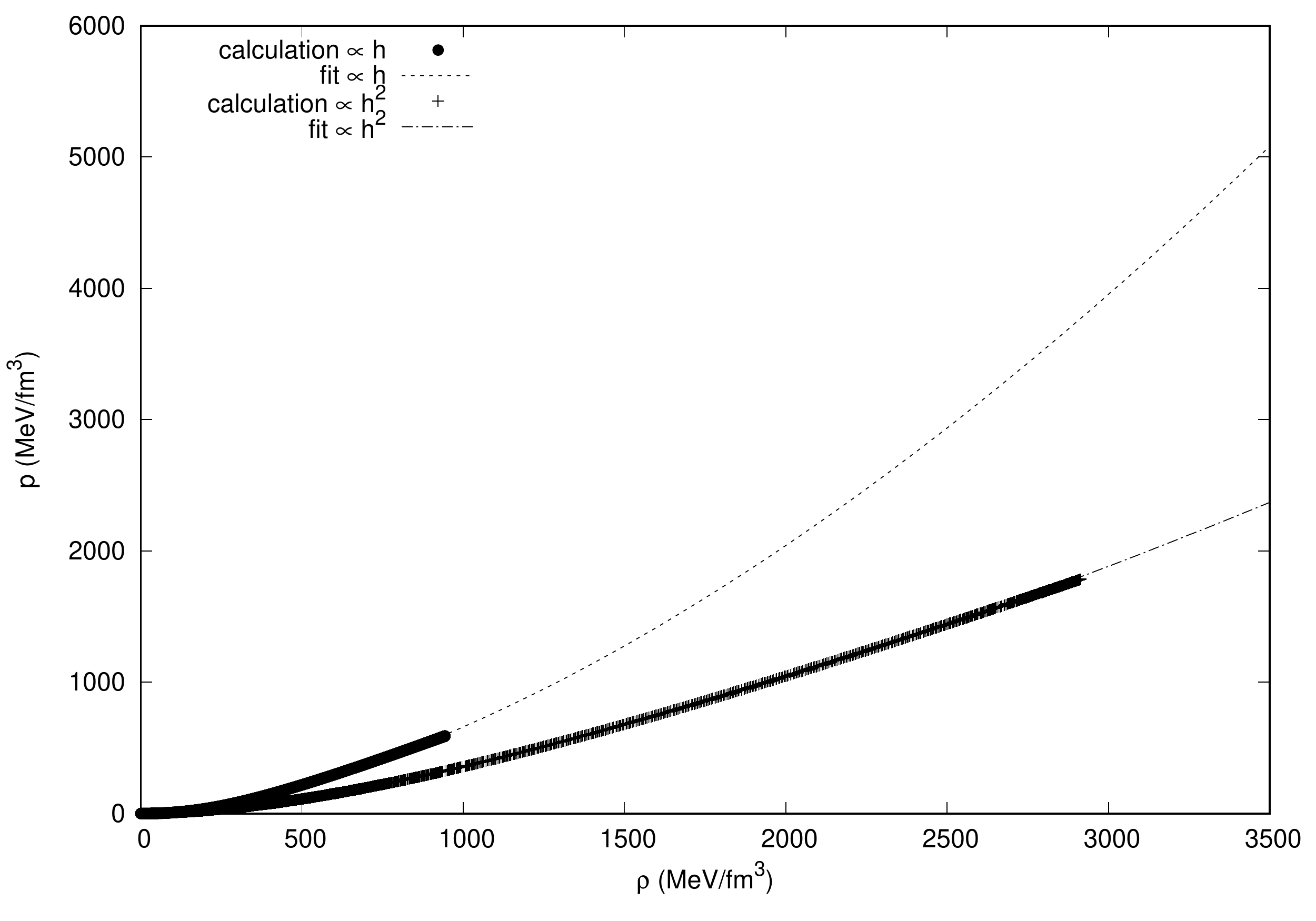}}
\caption{Equations of state of a polytropic type for the $\mathcal U_\pi$ (circle) and $\mathcal U_\pi^2$ (cross) potentials for baryon numbers $n=1$ (upper figure) and $n = n_{\rm max}$ (lower fugure). The lines show the corresponding fits. }\label{EoS}
\end{figure}

On the other hand, when focusing on their behavior at a local scale, bigger differences between the two approaches arise. For instance, as can be seen in Fig. \ref{EvsR}, the shape of the energy density as a function of the radial coordinate along the star depends on the chosen implementation. For the mean field, its value is almost constant, displaying a flat profile with a non-zero value at the surface of the star. Conversely, when the problem is solved in an exact way, the energy densities are clearly peaked around the center of the neutron star and vanishing at the surface. As previously commented, the shape of the pion mass potential squared makes this effect more pronounced than for the standard pion mass one.

Another case where this difference is evident corresponds to the metric field $\B$. For the mean field approach, the metric is a monotonically increasing field which reaches its maximum value at the star's surface. However, when the exact integration of Einstein's equations is performed, the maximum occurs in the interior of the star, further from the surface for the pion mass squared potential than for the standard pion mass (see Fig. 6 in Ref.~\refcite{Adam2015b}).

More interesting are the deviations between the two calculations when studying the equations of state. In the case of the mean-field, they are the same independently of the baryon number (or equivalently the mass) of the neutron star under consideration. On the contrary, this is not the behavior when the gravity back-reaction is included and the situation is subtler. Note that we do not have the energy density as a function of the pressure but both quantities in terms of the radial coordinate $r$, that is to say, $\rho = \rho(r)$ and $p = p(r)$. Nevertheless, we can eliminate this dependence from our numerical solutions and study the {\it on-shell} equations (the term on-shell is used here to remark that this has to be done once we have a particular solution). Interestingly, then the equations of state can be well approximated by a polytropic-like equation where the parameters depend now on the baryon number or total mass of the neutron star:
\begin{equation}
p = a(B) \rho^{b(B)},
\end{equation}

\noindent where we emphasize again that $a(B)$ and $b(B)$ are the fitting constants depending on the particular solutions.

In Fig. \ref{EoS} we show the on-shell equations of state and the corresponding polytropic curves for two different total masses of the pion mass and pion mass squared potentials. In each case we plot the EoS for the neutron stars with $n=1$ and $n=n_{\rm max}$, where $n = B/B_\odot$ is the baryon number in units of the sun baryon number defined as,
\begin{equation}
B_{\odot} = M_\odot/m_p = 1.188 \times 10^{57},
\end{equation}

\noindent (so it is not strictly speaking the baryon number of the sun) and $n_{\rm max}$ corresponds to the value when the maximal total mass has been achieved ($n_{\rm max} = 4.538$ and $n_{\rm max} = 2.963$ for $\mathcal U_\pi$ and $\mathcal U_\pi^2$, respectively). The parameters of the fit can be read off from Table \ref{PolyEq}.

\begin{table}[t] 
\caption{Parameter values from the on-shell EoS fits.}
\begin{center}
\begin{tabular}{@{}cccc@{}} \hline
Potential & $n=B/B_\odot$ & $a$ & $b$ \\ \hline
$\mathcal U_\pi = 2h$ & $1.0$ & $2.59 \times 10^{-5}$ & 2.34 \\
 & $4.538$ & $8.52 \times 10^{-3}$ & 1.63 \\
$\mathcal U_\pi^2 = 4 h^2$ & 1.0 & $5.17 \times 10^{-4}$ & 1.74  \\
 & $2.963$ & $1.21 \times 10^{-2}$ & 1.49 
\\ \hline \label{PolyEq}
\end{tabular}
\end{center}
\end{table}


\section{Conclusions}\label{Conclusions}

In this review we have given an insight in how the description of neutron stars within the Skyrme model has been following the same path as when topological solitons are used to describe nuclei. Hence, the first attempt corresponds to the use of the hedgehog ansatz. Unfortunately, the results in this case are unsatisfactory and the star is unstable with respect to single-particle decay for baryon number higher than one. However, this situation is not surprising since the same behavior appears when studying nuclei bigger than the nucleon.

Indeed, it is well known that for nuclei with baryon number greater than one the rational map ansatz is the right approximation. As we have discussed, a similar approach has been implemented for the study of neutron stars, where not only the rational map for large mass number has been considered but also a generalization to take into account multilayers where both the baryon number and shell width can even change along its radius. Nonetheless, such study cast a big problem, and although the radii of neutron stars are about the typical values, the masses are much below the massive neutron stars observed.

Since for large baryon number the true minimizer of the energy functional is a Skyrmionic crystal, the next step taken was to consider a cubic lattice of $\alpha$-particles. In this case, the solitonic configuration may be allowed to present some anisotropy by deviation from the fundamental cubic cell. The numerical simulations have shown that the system prefers an isotropic configuration but for massive stars close to the maximal mass allowed, where the anisotropy takes place. This implementation implies an improvement in neutron star masses with values close to but below $2 M_\odot$. Hovewer, the crystal nature of this description is at odds with the inner structure of neutron stars, where a perfect fluid-like behavior is expected.

To overcome these issues, the BPS Skyrme model has been recently proposed due to its perfect fluid attributes which are manifest in its energy-momentum tensor. In addition, due to the solvability property of the theory, a full numerical calculation with the gravity back-reaction included has been carried out besides the usual mean field approximation. These two different approaches allowed for comparison, showing that although the global properties of the system as masses or radii are of the same order, when looking at local properties as the energy densities or the metric fields, the descriptions are totally different. On the other hand, the stiffness of the corresponding equations of state allows for large star masses which are able to reach the high values observed in nature where other approaches fail.

All in all, the use of the BPS Skyrme model constitutes the most successful description of neutron stars using Skyrmions so far. Nonetheless, it is also known that some improvements are needed. One promising direction is to join the two main versions of the model reviewed here by introducing the terms coming from the standard Skyrme Lagrangian as a small contribution to the BPS model. In this way, we achieve a near-BPS theory which is thought to play a fundamental role not only for neutron stars but in nuclear physics in general. These additional terms are more important near the surface of the star (as opposed to the BPS model giving a bulk description) and, since the true minimizer of the energy for large $B$ is a lattice of Skyrmions, they will allow for a characterization of the crust where the state of matter is not a perfect fluid anymore but a lattice structure. Similarly, another possibility might be to couple the Skyrme model to one of the typical equations of state from nuclear physics, as well as introducing the neutron star rotation or the magnetic field. 

Finally, let us remark that the potential appearing in the BPS contribution may be arbitrary. The particular expressions chosen so far have been considered because of their role when applying the Skyrme model to nuclei. Notwithstanding, the study of different potentials is not only possible but desired, taking also into account the constraints imposed by observational data.


\section*{Acknowledgements}

The author is thankful to R. Vazquez for his help providing access to some data, suggestions and comments on the manuscript, and to C. Adam, J. Sanchez-Guillen and A. Wereszczynski for suggestions and comments on the manuscript. This work is supported by the INFN grant 19292/2017 {\it Integrable Models and Their Applications to Classical and Quantum Problems}.

\end{document}